\documentclass[useAMS,usenatbib,usegraphicx]{mn2e}
\title[RR Lyrae variables: visual and infrared luminosities, intrinsic colours, and kinematics]
{RR Lyrae variables: visual and infrared luminosities, intrinsic colours, and kinematics}
\author[Dambis et al.]{A. K. Dambis$^1$\thanks{E-mail: mirage@sai.msu.ru.}, L.N.~Berdnikov$^{1,5}$, A.Y.~Kniazev$^{1,2,3}$, V.V.~Kravtsov$^{1,4}$\\
\newauthor A.S.~Rastorguev$^1$, R.~Sefako$^2$, O.V.~Vozyakova$^1$\\ \\
$^1$ Sternberg Astronomical Institute, Lomonosov Moscow State University, Universitetskii pr. 13, Moscow, 119992 Russia\\
$^2$ South African Astronomical Observatory, P.O. Box 9, Observatory, Cape Town, 7935, South Africa\\
$^3$ Southern African Large Telescope, P.O. Box 9, Observatory, Cape Town, 7935, South Africa\\
$^4$ Instituto de Astronom$\acute{i}$a, Universidad~Cat$\acute{o}$lica~del~Norte, Avenida~Angamos, 0610, Antofagasta, Chile\\
$^5$ Isaac Newton Institute of Chile, Moscow Branch, Universitetskii Pr. 13, Moscow 119992, Russia}

\begin{document}

\date{}
\pubyear{2013}

\maketitle

\label{firstpage}

\begin{abstract}
We use UCAC4 proper motions and WISE $W1$-band apparent magnitudes intensity-mean for almost 400 field RR Lyrae variables
to determine the  parameters of the velocity distribution of Galactic RR Lyrae population and constrain the zero points
of the metallicity-$<M_V>$ relation and those of the period-metallicity-$<M_{K_s}>$-band  and 
period-metallicity-$<M_{W1}>$-band luminosity relations   via statistical parallax. We find the mean
velocities of the halo- and thick-disc RR Lyrae populations in the solar neighbourhood to be 
($U_0(Halo), V_0(Halo), W_0(Halo)$) =
$(-7 \pm 9, -214 \pm 10, -10 \pm 6)$ km s$^{-1}$ and  
($U_0(Disc), V_0(Disc), W_0(Disc)$) =
$(-13 \pm 7, -37 \pm 6, -17 \pm 4)$ km s$^{-1}$, respectively, and the 
corresponding components of the velocity-dispersion ellipsoids, 
($\sigma V_R(Halo), \sigma V_{\phi}(Halo), \sigma V_{\theta}(Halo)$) =
$(153 \pm 9, 101 \pm 6, 96 \pm 5)$ km s$^{-1}$ and  
($\sigma V_R(Disc), \sigma V_{\phi}(Disc), \sigma V_{\theta}(Disc)$) =
$(46 \pm 7, 37 \pm 5, 27 \pm 4)$ km s$^{-1}$, respectively. The fraction of
thick-disc stars is estimated at 0.22 $\pm$ 0.03. 
The corrected IR period-metallicity-luminosity relations are 
$<M_{K_s}>$ = -0.769 +0.088 $\cdot$ [Fe/H]- 2.33 $\cdot \mathop{\rm log} P_F$ 
and $<M_{W1}>$ = -0.825 + 0.088$\cdot$ [Fe/H] -2.33 $\cdot \mathop{\rm log} P_F$,
and the optical metallicity-luminosity relation, 
[Fe/H]-$<M_V>$,  is $<M_V>$ = +1.094 + 0.232$\cdot$ [Fe/H],
with a standard error of $\pm$ 0.089,
implying an LMC distance modulus of 18.32 $\pm$ 0.09, a solar 
Galactocentric distance of 7.73 $\pm$ 0.36~kpc, and the M31 and M33 distance
moduli of $DM_{M31}$ = 24.24 $\pm$ 0.09 ($D$ = 705 $\pm$ 30 kpc)
and $DM_{M33}$ = 24.36 $\pm$  0.09 ($D$ = 745 $\pm$ 31 kpc), respectively. 
Extragalactic distances calibrated with our RR Lyrae star luminosity scale
imply a Hubble constant of $\sim$80~km/s/Mpc.
Our results suggest marginal prograde rotation for the 
population of halo RR Lyraes in the Milky Way. 

\end{abstract}

\section[1]{Introduction}

RR Lyrae variables are A-F type giants undergoing core helium burning, which pulsate with periods 0.2-1.2~d and have masses 
and luminosities of $M \sim 0.7 M_{\odot}$ and  $L = 40-50 L_{\odot}$, respectively. 
They  are easy to identify by their periods and characteristic
light-curve shapes and, like other pulsating stars, obey a period-metallicity-luminosity relation, 
which in the case of the $V$- and $K$- bands appears to degenerate into the metallicity-luminosity 
and period-luminosity relations, respectively \citep{catelan}, both characterised by a very low
intrinsic scatter of $\leq 0.06-0.09^m$ \citep{wks,fs}. Because of the large
ages ($>$~10~Gyr) of RR Lyrae variables these stars occur in systems with old populations (globular clusters,
elliptical galaxies, halos and thick discs of spiral galaxies, irregular galaxies) and can therefore serve 
as distance indicators, as well as kinematic and metallicity tracers. Although their factor of $\sim$~100 lower luminosities have 
long left RR Lyraes somewhat overshadowed by Cepheids as extragalactic distance indicators, 
recent progress in extragalactic stellar photometry has made them an efficient tool for mapping the Local Group, and 
one of the most important contributors to the first rung of the cosmic distance scale \citep{cacciari}.

However, for RR Lyraes to be properly usable to this end, the parameters of the
period-metallicity-luminosity relations in various bands have to be established with adequate precision.
The problem is that although a consensus seems to have emerged concerning the slopes of the 
corresponding relations in various photometric bands, this still is
far from true for the respective zero-points: they, like 17 years ago, "continue to defy 
a consensus" \citep{l1} with estimates reported by different authors spanning a $\sim$~0.3-0.5$^m$
wide interval.

Six decades ago, Pavlovskaya (1953) was the
first to apply the statistical-parallax technique to estimate the
mean absolute magnitude of RR Lyrae stars. Since then, 
the method, continuously refined, has been used extensively by many authors
\citep{p1,r1,vh,cd1,h2,str,l1,pg1,pg2,gp,f1,ts,luri,dr1,d1,dv,rdz, dambis2,k12} with
ever increasing samples of RR Lyrae stars.
The current, maximum-likelihood version of the technique was first
proposed by \citet{m1}, and its practical application dates back to
\citet{str} and \citet{h2}. \citet{l1} were the first to take into account 
the kinematic inhomogeneity of the local RR Lyrae population by a priori attributing each star
either to the halo or thick-disc subsample based on metallicity and velocity data. 
\citet{lm} generalised
the method to explicitly incorporate the eventual multicomponent
structure of the population studied, and \citet{luri} applied the generalised technique 
to Galactic RR Lyrae type stars. One of the solutions in our previous study \citep{dambis2} 
also took  into account the bimodal nature of the velocity distribution
without a priori partitioning the sample into two groups.

Time has now come for yet another statistical-parallax based study of RR Lyrae type stars to be performed
for two main reasons. First, impressive progress has been achieved in the photometry of RR Lyrae variables, 
with good light curves obtained for many stars. These include (a)  extensive optical data
acquired both within the framework of ASAS 
survey~\citep{pojmanski} and as a result of our dedicated program of photometric observations of 
RR Lyrae stars \citep{ber1, ber2, ber3}, and (b) infrared light curves acquired within the framework of
WISE all-sky photometric survey \citep{wise} supplemented by single-phase
2MASS $JHK_s$-band measurements \citep{twomass}, which, given precise light elements,
can be converted to the corresponding intensity-mean magnitudes \citep{flkvw}. As a result,
bona fide multicolour intensity-mean magnitudes have for the first time become available simultaneously 
for the bulk of $\sim$~400  Galactic field RR Lyrae type variables with known
radial velocities and metallicities, allowing the interstellar extinction to be accurately
determined for all these objects. Second,
the release of the final, fourth version of  The US Naval Observatory CCD Astrograph Catalogue
(UCAC4)\citep{uc4}, which provides very accurate proper motions for  most of
the stars down to an $R$-band limiting magnitude of about 16$^m$, 
allows the statistical-parallax study to be based on a single source of proper motions
throughout the entire sky.

The primary aim of this paper is to refine the infrared and visual
photometric distance scales of RR Lyrae variables by further constraining 
the zero points of the $V$-, $K_s$-, and WISE $W1$-band period-metallicity-colour relations 
for these stars via the method of statistical parallax, and use them to estimate the distances to
a number of Local group galaxies. As a by product,
we calibrate the $(V - K_s)_0$ and $(V - W1)_0$ intrinsic colours of these stars 
and determine the kinematical parameters of the local RR Lyrae star population.

Section~2 describes the observational data that we use for our statistical-parallax analysis.
In Section~3 we analyze the interstellar extinction toward our sample stars and derive
the intrinsic-colour relations for RR Lyrae type stars in terms of fundamental period
and metallicity. Section~4 is devoted to the initial (provisional) distances to our sample
RR Lyraes to be adjusted. In Section~5 we say a few words about the statistical-parallax method employed
and the (kinematical and distance-scale) parameters to be determined.
In Section~6 we present the results of our analysis.
In Sections~7, 8, and 9  we discuss the implications of our results for the cosmic distance scale,
rotation of RR Lyrae populations, and cosmology, respectively. 
The final section summarizes  the conclusions.

\section[]{OBSERVATIONAL DATA}

First, we tried to collect the most complete possible sample of Galactic field
RR Lyrae type variables with measured radial velocities and metallicities. The sample
we use in this paper is based mostly on that employed in our previous statistical-parallax
analysis of RR Lyrae variables \citep{dambis2}, which, in turn, is based on the list of
Galactic RR Lyraes from the revised catalogue of 2106 Galactic stars selected
without kinematic bias and with available radial velocities,
distance estimates, and metal abundances in the range
-4.0$\leq$[Fe/H]$\leq$0.0 compiled by \citet{b1}. This list
includes a total of 388 Galactic RR Lyrae stars, which we supplement with the data for 14 additional RR Lyrae 
type variables from recent papers. Thus our final sample contains 402 RR Lyrae type stars.

\subsection{Periods and pulsation modes}

Like in our previous study \citep{dambis2}, we adopt 
the periods  of the RR Lyrae type stars of our sample from the 
ASAS catalogue \citep{pojmanski}, \citet{maintz}, or the General Catalogue of Variable Stars
\citep{gc1}. The latter was our source of pulsation modes. We use the 
formula $P_F$ = log $P$ + 0.127 \citep{fs} to fundamentalise the periods of
RRc type variables (first-overtone pulsators).

\subsection{Apparent intensity-mean magnitudes}

\subsubsection{$V$-band data}

\begin{figure}
\includegraphics[width=0.5\textwidth]{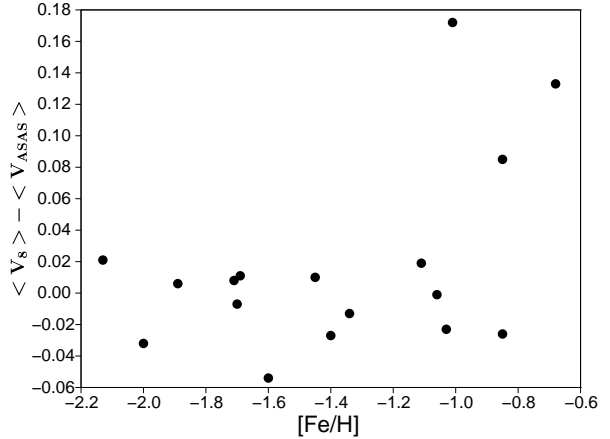}
  \caption{Intensity-mean magnitude difference between dataset (8) \citep{l2, l4, l5} and ASAS as a 
function of metallicity.}
 \label{feh_systematics}
\end{figure}

\citet{l1} characterised the available $V$-band photometry for Galactic field RR Lyrae variables 
as a "surprisingly heterogeneous data set", and this is still true after 17 years, although 
extensive optical data has been acquired since then as a result of ASAS 
survey~\citep{pojmanski} and our own dedicated program  \citep{ber1, ber2, ber3}.
In this paper, we derive the intensity-mean Johnson $V$-band magnitudes $<V>$ 
based on nine sufficiently large overlapping
sets of photoelectric and CCD observations (\citet{bookmeyer} (1), ASAS \citep{pojmanski} (2),
HIPPARCOS $H_p$-band photometry \citep{hi} (3), \citet{lub, vgb} (4), \citet{cd1} (5),
our own data \citet{ber1, ber2, ber3}  (6), \citet{schmidt, schmidt2, schmidt3} (7),
\citet{l2, l4, l5} (8), \citet{kinman1, kinman3, kinman2, kinman4} (9)). 
In the case of the data of \citet{bookmeyer} and \citet{cd1} we computed
the initial intensity means $<V>$ using Eq.~(1) from \citet{l2}. We adopted the initial 
intensity-mean $<V>$ magnitudes based on HIPPARCOS $H_p$-band
photometry from \citet{f1} or computed them in accordance with the procedure described by the
above authors for stars lacking in their paper. In the case of observations made in the
five-colour Walraven system we adopted the Johnson magnitude means $<V>_m$ listed by \citet{lub}
and computed the corresponding magnitude means based on the data reported by \citet{vgb}. In the
case of ASAS data we determined the intensity means $<V>$ from the light-curve fits
defined by the Fourier coefficients reported by \citet{Szczygiel} or computed them
directly from ASAS data for the stars lacking in the list of \citet{Szczygiel}. 
We thus have a set of mean magnitudes $<V_{ik}>$ for our stars, where $i$ and $j$
are the number of the dataset and the number of the star, respectively. As mentioned above,
$<V_{ik}>$ is the initial intensity-mean magnitude estimate, except for the case
of dataset $i$=4 \citep{lub, vgb}, which consists of magnitude means. 
We then determine the homogeneous intensity-mean $<V_{k}>_{final}$ magnitudes 
for our stars and the 
correcting offsets $\Delta V_i$ by solving the following set of linear equations:
\begin{equation}
<V_{k}>_{final} - \Delta V_i = <V_{ik}>  
\end{equation}
We adopt the dataset of \citet{bookmeyer} as our standard system and set
$\Delta V_1$ = 0 by definition. We collected a total of 905 mean magnitude estimates
for 384 stars and hence have 905 equations for 384+8=392 unknowns.
The resulting offsets $\Delta V_i$ are listed in column~4 of Table~\ref{Vsource}. Other
columns list the number of the dataset (column~1), the references to the original data (column~2), 
the number of stars in the dataset (column~3), and remarks (column~5).  The standard errors of the
homogenised intensity-mean magnitudes $<V_{k}>_{final}$ range from $\sim$~0.015$^m$ to
$\sim$~0.035$^m$. The offsets can be
seen to be generally small, and only in three cases do they appreciably exceed 0.01$^m$ in absolute value:
these correspond to datasets (4) \citep{lub, vgb}, (5) \citep{cd1}, and (7)\citep{schmidt, schmidt2, schmidt3}. 
To reveal eventual metallicity-dependent systematics, we computed the intensity-mean magnitude differences 
$<V_{ik}> - <V_{2,k}>$ between each dataset and the largest dataset~(2)~(ASAS) and fitted them to a linear
relation of the form
\begin{equation}
<V_{ik}> - <V_{2,k}> = p_i + q_i \cdot [Fe/H].
\end{equation}
We found, as expected, the resulting metallicity slopes $q_i$ not to differ significantly from zero and range from
-0.002~$\pm$0.023 to +0.020~$\pm$~0.046 for all datasets except no.~8 \citep{l2, l4, l5}:
$$
<V_8> - <V_{ASAS}> = 
$$
\begin{equation}
+0.104~\pm~0.045 + (0.063~\pm~0.031) \cdot [Fe/H].
\end{equation}
However, even in this case the slope barely exceeds 2$\sigma$ and is actually due to three clear outliers at the
high-metallicity end (see Fig.~\ref{feh_systematics}). Note that two of these stars --- UY~Aps ($b = -11.39$) and V494~Sco 
($b = -0.49$) --- are located in crowded fields close to the Galactic equator, where ASAS performance is degraded because
of the large pixel size ($\sim~15~arcsec^2$). We therefore homogenize $V$-intensity magnitudes without introducing any
metallicity dependent terms. 

\begin{table*}
  \centering
  \begin{minipage}{140mm}
  \caption{Sources of intensity-mean $V$-band magnitudes (based on photoelectric and CCD photometry)}\label{Vsource}
\begin{tabular}{r  l r r r}
\hline
  Dataset &  References & Number & $\Delta V_i$ & Remark\\
no. (i) &  & of stars & & \\
  \hline
1 &  \citet{bookmeyer} & 142 & 0.000~$\pm$~0.000 & \\
\hline
2 & ASAS \citep{pojmanski} & 234 & -0.001~$\pm$~0.004 & \\
\hline
3 & HIPPARCOS \citep{hi,f1} & 148 & +0.009~$\pm$~0.004 & A \\
\hline
4 &  \citet{lub, vgb} & 83 & +0.054~$\pm$~0.005 & B \\
\hline
5 &  \citet{cd1} & 59 & -0.040~$\pm$~0.006 &\\
\hline
6 & \citet{ber1, ber2, ber3} & 120 & -0.005~$\pm$~0.005 &  \\
\hline
7 & \citet{schmidt, schmidt2} & 57 & +0.018~$\pm$~0.007 & \\
  & \citet{schmidt3} & & \\
\hline
8 & \citet{l2, l4, l5} & 23 & +0.013~$\pm$~0.010 & \\
\hline
9 & \citet{kinman1, kinman3, kinman2} & 28 &  +0.014~$\pm$~0.013 &\\
   & \citet{kinman4} & &\\
\hline
Remarks: {\bf A}. & Johnson's $V$-band intensity-mean magnitudes & &\\
         & converted from HIPPARCOS $<Hp>$ magnitudes  & & \\
         & as described by \citet{f1}& & \\
         {\bf B}. & Based on magnitude means in the  Walraven   & & \\
         & five-colour VBLUW system. Johnson's $V_J$-band    & & \\
         & magnitude means computed by the formula  & & \\
         &  $V_J$ = 6.874$^m$ - 2.5 ${V  + 0.065(V-B)}$  \citep{pel} & & \\
\end{tabular}
\end{minipage}
\end{table*}

\begin{figure}
\includegraphics[width=0.5\textwidth]{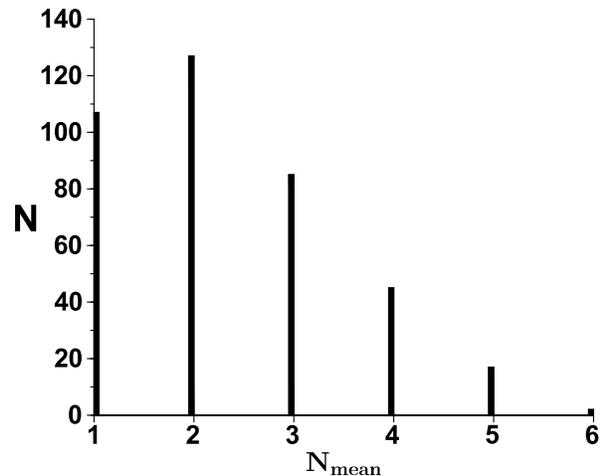}
  \caption{Histogram of the number of $<V>$ sources per star.}
 \label{nv}
\end{figure}

\begin{figure}
\includegraphics[width=0.5\textwidth]{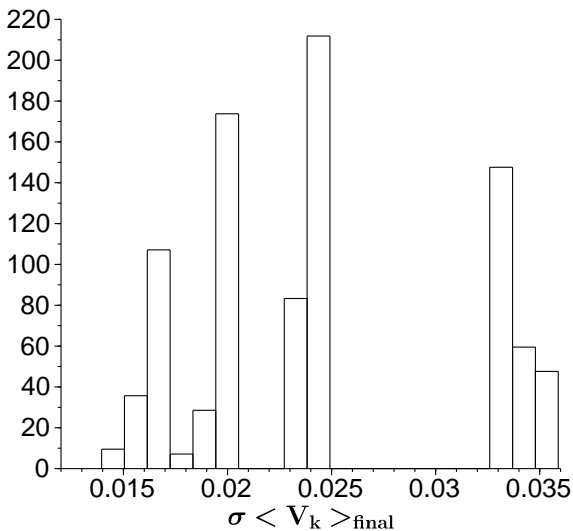}
  \caption{Histogram of standard errors of the final intensity-mean $<V>$-band 
magnitudes.}
 \label{sigv}
\end{figure}

Figure~\ref{nv} shows the distribution of the stars by the number of mean magnitude estimates
and Fig.~\ref{sigv}, the distribution of the standard errors of the final $V$-band intensity means,
$\sigma <V_{k}>_{final}$.

\subsubsection{$K_s$-band data}

As we already pointed out \citep{dambis2}, the 2MASS project, which mapped the entire sky in three
near-infrared bands -- $J$, $H$, and $K_s$, provides homogeneous near-infrared photometry for almost all
Galactic RR Lyrae type variables with known metallicities and
radial velocities. Another important advantage of 2MASS data over available optical photometry is
its highly reduced sensitivity to interstellar extinction, which is one order of magnitude smaller
in the $K_s$ band than in the $V$ band \citep{yuan}. 
We adopt the intensity-mean $K_s$-band magnitudes for most of the stars of our sample from 
\citet{dambis2}. These are  single-epoch 2MASS magnitudes phase corrected in accordance with the procedure 
described by \citet{flkvw} (including the data adopted from \citet{kinman} for 13 stars), 
which are accurate to within $\sim 0.03^m$ (summed quadratically
with error of the observed 2MASS $K_s$ magnitude), or just raw 2MASS magnitudes with no phase
correction applied (in the case of no or too outdated ephemeris) for
32 stars. For the additional 12 stars 
we determine the intensity-mean $K_s$-band magnitudes by phase correcting their 2MASS $K_s$-band magnitudes
using the same procedure described by \citet{flkvw} (for five stars) or adopting  the phase-corrected 2MASS $K_s$ magnitude
from \citet{kinman} (one star) and \citet{kinman5} (six stars). We have 
collected intensity-mean $K_s$-band estimates
for all 403 RR Lyrae type variables of our sample.

\subsubsection{WISE $W1$-band data}

\begin{figure}
\includegraphics[width=0.5\textwidth]{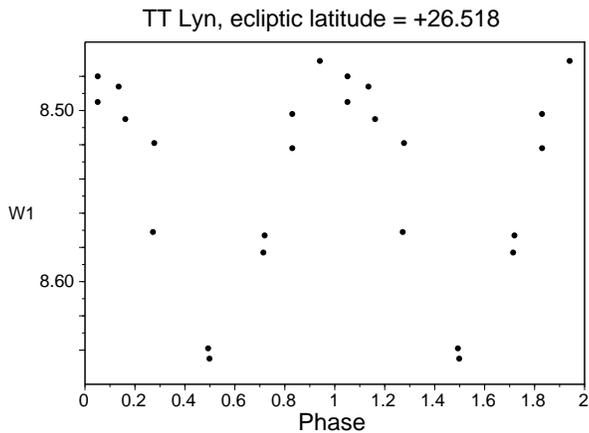}
  \caption{WISE W1 light curve of TT Lyn.}
 \label{lcwise1}
\end{figure}

\begin{figure}
\includegraphics[width=0.5\textwidth]{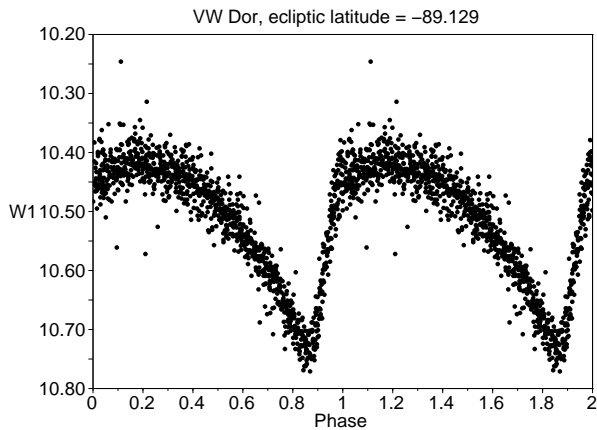}
  \caption{WISE W1 light curve of VW Dor.}
 \label{lcwise2}
\end{figure}

The WISE All-Sky Data Release, which was made public last
year \citep{cutri_wise} and which mapped the entire sky in four
mid-infrared bands W1, W2, W3, and W4 with the effective wavelengths 
of 3.368, 4.618, 12.082 and 22.194~$\mu$m, respectively \citep{wise},
is even better suited for building the distance scale of RR Lyrae type variables.
Its chief advantages over 2MASS are the true homogeneity (measurements performed 
over the entire sky with a single instrument operating beyond the earth atmosphere),
a factor of $\sim$~1.7 smaller sensitivity to interstellar extinction \citep{yuan},
and good phase coverage for short-period variables
with at least a dozen well-distributed measurements available for all stars and 
over a thousand measurements made for stars near the ecliptic poles. We used
the WISE single-exposure database to compute the intensity-mean average W1-band 
magnitudes, $<W1>$, for a total of 398 stars of our sample. Figure~\ref{lcwise1} 
shows the WISE W1 light curves for a TT Lyn, a typical RR Lyrae variable from our list,
and Fig.~\ref{lcwise2}  shows the WISE W1 light curve of VW Dor, which is located
close to the South Ecliptic Pole at the ecliptic latitude of -89.129$^{\circ}$. 

\subsection{Metallicities}

\begin{figure}
\includegraphics[width=0.5\textwidth]{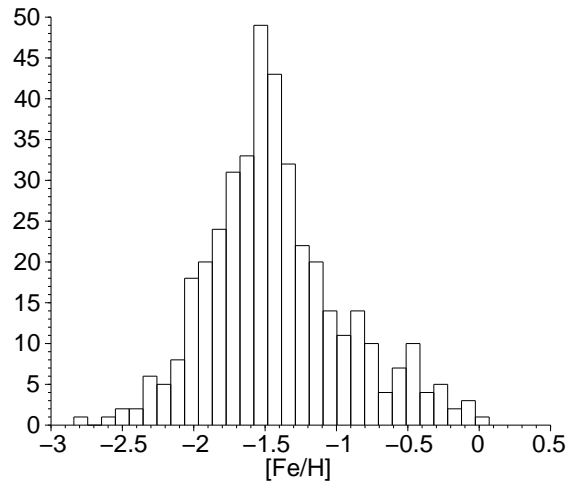}
  \caption{Distribution of metallicities of sample stars.}
 \label{fehdist}
\end{figure}

For most of the stars of our sample we use the homogenised metallicities on
the \citet{zw}  scale adopted from the compilation of
\citet{dambis2},  which, in turn, is mostly based on the catalogue of \citet{b1}. All these
metallicities are based on spectroscopic measurements. Our sources of metallicities 
for the additional stars  lacking in the sample of \citet{dambis2} were \citet{fb} 
(BH~Aur, V363~Cas, and EZ~Cep, reduced to the Zinn and West scale via formula~(46) in \citet{dambis2}), 
\citet{l2} (ET~Hya, , already on the Zinn and West scale), 
\citet{p2} (RW~Lyn, reduced to the Zinn and West scale via formula~(46) in \citet{dambis2}), \citet{kinman}
(MO~Com, reduced to the Zinn and West scale via formula~(46) in \citet{dambis2}), 
\citet{rotse} (EN~Lyn, photometric metallicity on the Zinn and West scale), \citet{kemper} (BK~UMa and BN~UMa,
delta S values reduced to [Fe/H] on the Zinn and West scale).
We computed the photometric metallicities of CH~Aql (from ASAS light-curve parameters) 
and DQ~Lyn (from NSVS light-curve parameters) using 
formula~(14) from \citet{rotse} and formula~(3) from \citet{morgan}, respectively.
We thus have collected homogenised metallicities for a total of
402 stars of our sample (no metallicity could be determined for CK~UMa). 
Figure~\ref{fehdist} shows the distribution of metallicities for stars
of our sample.

\subsection{Radial velocities}

\begin{figure}
\includegraphics[width=0.5\textwidth]{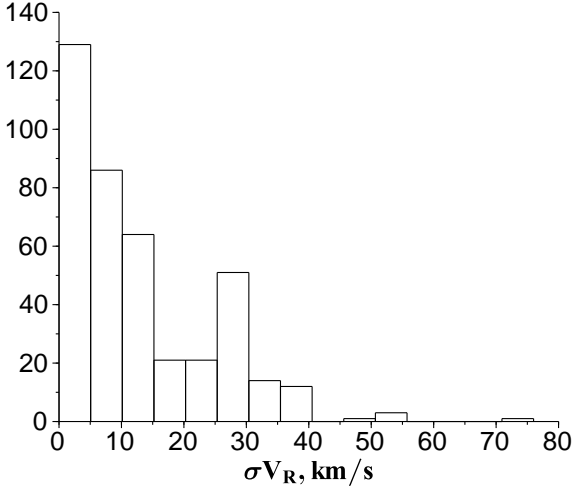}
  \caption{Distribution of radial-velocity errors of sample stars.}
 \label{sigvr}
\end{figure}

We adopted the mean radial
velocities, $V_R$, and their standard errors, $\sigma V_R$, for most of the stars from
\citet{dambis2}, and the corresponding quantities for the additional
stars, from \citet{fb} (BH~Aur, V363~Cas, and EZ~Cep), \citet{j2} (CH~Aql and BK~UMa), \citet{p2} (RW~Lyn), \citet{kinman} (MO~Com),
\citet{kinman4} (DQ~Lyn, EN~Lyn, BN~UMa, and CK~UMa).
We also updated the $\gamma$ velocities for seven stars (WY~Ant, XZ~Aps, BS~Aps, Z~Mic,
RV~Oct, V1645~Sgr, and AS~Vir) based on the data reported by \citet{for}.
We collected the mean radial velocity 
estimates  for all 403 stars of our sample.
Figure~\ref{sigvr} shows the distribution of radial-velocity errors for stars
of our sample.

\subsection{Proper motions}

\begin{figure}
\includegraphics[width=0.5\textwidth]{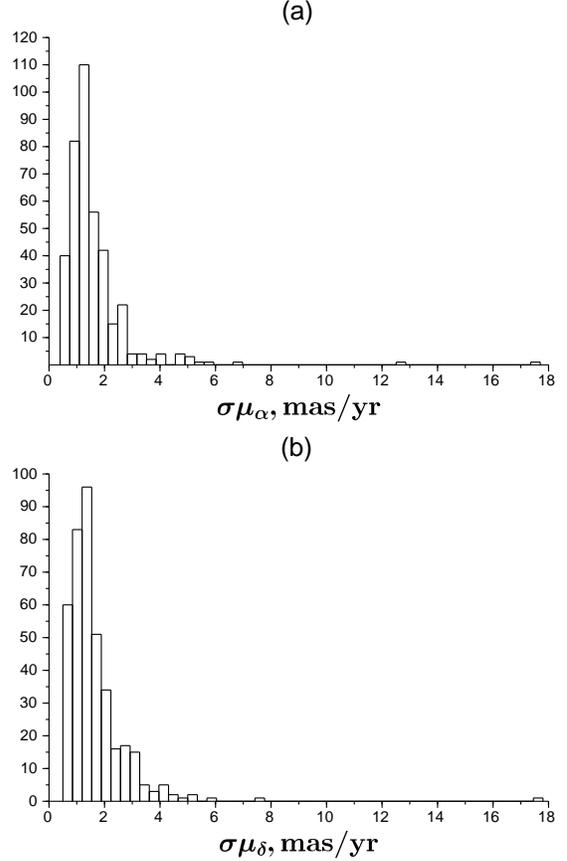}
  \caption{Distribution of proper-motion errors of sample stars
in right ascension (a) and declination (b).}
 \label{sigpm}
\end{figure}

We adopt the fourth United States Naval Observatory (USNO) CCD Astrograph Catalogue, UCAC4,
\citep{uc4}, which "is complete from the brightest stars to about magnitude R = 16",
as our only source of absolute proper motions, which are available for
393 of the 403 stars of our list.
Figure~\ref{sigpm} shows the distribution of proper-motion errors in right ascension (the top panel) and
declination (the bottom panel).

\section{Interstellar extinction, period-metallicity-luminosity relations,
and intrinsic colour calibrations}

We now estimate  the amount of interstellar extinction toward stars of our sample. The most straightforward
way is to use a 2D extinction map and assign to each star the "extinction at infinity". This approach should 
work well for most of the stars except those located close to the Galactic plane. We consider it better to compute
extinction in terms of the 3D extinction model by \citet{drimmel}. To do this, we need to approximately
estimate the distances to our stars (we show below that choosing another, even significantly different published relation,
has no effect on the colour calibrations and extinction estimates obtained in this section). 
We begin by estimating the absolute $K_s$-band magnitudes of our sample stars by
adopting (provisionally)  the following period-luminosity
relation derived by \citet{j1} based on the results obtained using the Baade-Wesselink method:
\begin{equation}
 <M_{K(Jones)}> = -2.33 \cdot \mathop{\rm log} P_F - 0.88,
\label{MKJ}
\end{equation}
where $K$-band magnitudes are on the CIT system. Given the negligible and statistically
insignificant coefficient of the colour term in the transformation equation
(http://www.astro.caltech.edu/~jmc/2mass/v3/transformations/)

$$
(K_s)_{2MASS} = K_{CIT} + 
$$

\begin{equation}
(-0.019 \pm 0.004) + (0.001 \pm 0.005)(J-K)_{CIT}
\end{equation}

and the fact that average $(J-K)$ colour indices of RR Lyrae type variables always
lie inside the interval from 0.00$^m$ to 0.50$^m$, period-$K_{CIT}$-band luminosity relation given by equation~(\ref{MKJ})
transforms into the following period-$K_s$-band luminosity relation
\begin{equation}
 <M_{K_s}> = -2.33 \cdot \mathop{\rm log} P_F - 0.90,
\label{MKs}
\end{equation}
which we use to compute the absolute $K_s$-band magnitudes of our stars.  We then use the 3D extinction 
model by \citet{drimmel} to determine the amount of interstellar extinction toward each star via  
the following iterative procedure. We initially set
the total $K_s$ band extinction  equal to zero, $A_{K_s}$=0, and compute the distance (in kpc) by the following formula:
\begin{equation}
d = 10^{0.2\cdot(<K_s>-A_{K_s} - M_{K_s})-2}.
\label{iter}
\end{equation}
We use this distance estimate and the star's coordinates to compute the refined estimate of the amount of interstellar
extinction, $A_V$,  in terms of the model of \citet{drimmel}, transform it into $E(B-V)$ = $A_V/R_V$ ($R_V$ = 3.1 \citep{yuan}), 
compute $A_{K_s}$ = $R_K\cdot E(B-V)$ ($R_{K_s}$ = $A_{K_s}/E(B-V)$ = 0.306 \citep{yuan}), and refine the distance estimate via formula~(\ref{iter}).
We repeat this procedure until both $d$ and $A_{K_s}$ (and, naturally, $E(B-V)$, and $A_V$) converge.

A question naturally arises to which extent are our reddening estimates dependent on the adopted provisional
distances to RR Lyrae type stars. To assess the effect of a different set of input distance estimates, we repeated the
above procedure using two other popular period-metallicity-$<M_K>$ relations, which, compared to PL relation~(\ref{MKs}), 
yield brighter absolute magnitudes practically for all our stars,
on the average by $\sim$~0.35$^m$, translating into a factor of $\sim$~1.17 longer distances. The one was derived theoretically by \citet{bono}:
$$
<M_{K_s}> =  -2.101 \cdot \mathop{\rm log} P_F +0.231 (\pm~0.012) \cdot [Fe/H]
$$
\begin{equation}
  -0.770 (\pm~0.044),
\label{MKBono}
\end{equation}
and the other is an empirical one derived by \citet{sollima} based on observational data for globular-cluster RR Lyrae variables:
$$
<M_{K_s}> =  -2.38 (\pm~0.04) \cdot \mathop{\rm log} P_F +0.09 (\pm~0.14) \cdot [Fe/H]
$$
\begin{equation}
 -1.07 (\pm~0.11),
\label{MKSollima}
\end{equation}
We found the distances based on the period-metallicity-$<M_K>$ relation of \citet{bono} to yield $A_V$ values that
differ by +0.0023$\pm$0.0004 with a scatter of 0.009 from those computed with distances based on relation~(\ref{MKs}). The 
corresponding difference for the calibration of \citet{sollima} is +0.0031$\pm$0.0006 with a scatter of 0.011.
The differences become truly negligible for the subsample of stars with $|b|~>~25^{\circ}$, on which our final 
colour calibrations are actually based (see below):  +0.0002$\pm$0.0001 with a scatter of 0.0016 for both the 
relation of \citet{bono} and that of \citet{sollima}. We can therefore safely assume that all the colour calibrations derived below
do not depend on the provisional input distances used to estimate the interstellar extinction toward the calibrating 
RR Lyrae stars of our sample. Hence the final colour calibrations remain unchanged and so are the extinction values for
low-latitude stars that are based on them. We point out that the provisional distances computed in this section serve  the
sole purpose of estimating the interstellar extinction and are not used below for any other end.

We now use the resulting $A_V$, $A_{K_s}$, and $A_{W1} = R_{W1} \cdot E(B-V)$ ($R_{W1}$ = $A_{W1}/E(B-V)$ = 0.18 
\citep{yuan}) values to deredden the observed
intensity-mean $<V>$, $<K_s>$, and $<W1>$ magnitudes:
\begin{equation}
<V>_0 = <V> - A_V, 
\label{DRV}
\end{equation}

\begin{equation}
<K_s>_0 = <K_s> - A_{K_s}, 
\label{DRK}
\end{equation}

\begin{equation}
<W1>_0 = <W1> - A_{W1},
\label{DRW}
\end{equation}

and the intrinsic colours:

\begin{equation}
(<V>-<K_s>)_0 = <V> - <K_s> - ( A_V - A_{K_s}), 
\label{VK}
\end{equation}

\begin{equation}
(<K_s> - <W1>)_0 = <K_s> - <W1> - (A_{K_s} - A_{W1}), 
\label{KW}
\end{equation}

\begin{equation}
(<V> - <W1>)_0 = <V> - <W1> - (A_V - A_{W1}).
\label{VW}
\end{equation}

We now calibrate the $(<V> - <K_s>)_0$ intrinsic colour index
in terms of [Fe/H] and log $P_F$. We proceed based on the following two established facts.
First, the absolute V-band magnitude of RR Lyrae variables depends on metallicity [Fe/H] and,
for a given metallicity is independent of period \citep{catelan}:
\begin{equation}
<M_V> = a_v + b_v \cdot [Fe/H],
\label{MV}
\end{equation}
where the most recent reliable direct empirical estimates of the slope $b_v$ are $b_v$~=~0.214~$\pm$~0.047
\citep{gratton} and $b_v$~=~0.25~$\pm$~0.02 \citep{federici}. These estimates are based on the sole 
geometric assumption that the stars in question are in both cases at the same distance from us. 
Hereafter we adopt the simple (unweighted)
average of the two:
\begin{equation}
<M_V> = a_v + 0.232 (\pm~0.020) \cdot [Fe/H],
\label{MVF}
\end{equation}
We prefer to rely on the above estimates rather than on those based 
on Baade-Wesselink analyses \citep{cacciari92, skillen, f2}, which depend on more input assumptions and 
are therefore not enough "empiric". However, we point out that the corresponding slope estimates agree well with those
mentioned above:  $b_v$~=~0.20~ \citep{cacciari92},  $b_v$~=~0.21~$\pm$~0.05 \citep{skillen}, 
and $b_v$~=~0.20~$\pm$~0.04 \citep{f2}.
Second, the absolute $K_s$-band magnitude of RR Lyrae variables depends on fundamentalised pulsation period, $P_F$,
and, possibly, metallicity
\begin{equation}
<M_{K_s}> = a_k + b_k \cdot [Fe/H] + c_k \cdot log P_F,
\label{MK}
\end{equation}
with the $log P_F$ slope of $c_k$ = -2.33 \citep{j1, fs, catelan}, i.e.:
\begin{equation}
<M_{K_s}> = a_k + b_k \cdot [Fe/H] -2.33 \cdot log P_F.
\label{MK1}
\end{equation}

\begin{figure}
\includegraphics[width=0.5\textwidth]{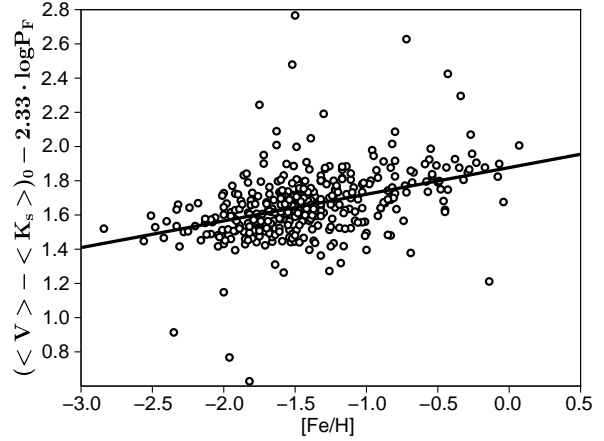}
  \caption{Plot of $(<V>-<K_s>)_0 - 2.33 \cdot log P_F$ as a function of [Fe/H]. The
solid line shows linear fit (equation~(\ref{abvk})) with parameters (\ref{abvkpar}).}
 \label{figvk}
\end{figure}

We subtract equation~(\ref{MK1}) from equation~(\ref{MV}) to obtain:
$$
(<V>-<K_s>)_0 = <M_V> - <M_{K_s}> = 
$$
\begin{equation}
= (a_v - a_k) + (b_v - b_k) \cdot [Fe/H] + 2.33 \cdot log P_F.
\end{equation}
We denote $a_{vk} = a_v - a_k$ and $b_{vk} = b_v - b_k$ to rewrite the above equation as:
\begin{equation}
(<V>-<K_s>)_0  = a_{vk} + b_{vk} \cdot [Fe/H] + 2.33 \cdot log P_F,
\label{abvk}
\end{equation}
where the only two unknown quantities are $a_{vk}$ and $b_{vk}$.
We use the linear least squares technique (with 3$\sigma$ clipping) to 
solve the set of equations~(\ref{abvk}) for all 383 stars with available
$<V>$, $<K_s>$, and [Fe/H] to find:

\begin{eqnarray} \label{abvkpar}
(<V>-<K_s>)_0  = (1.883 \pm 0.019)   \\  \nonumber
+ (0.159 \pm 0.013) \cdot [Fe/H] + 2.33 \cdot log P_F,  \\  \nonumber
\end{eqnarray}
with a scatter of $\sigma(<V>-<K_s>)_0$ = 0.12.
Figure~\ref{figvk} plots the left-hand side of equation~(\ref{abvk}) as a function [Fe/H]
with linear fit with parameters~(\ref{abvkpar}) superimposed.

\begin{figure}
\includegraphics[width=0.5\textwidth]{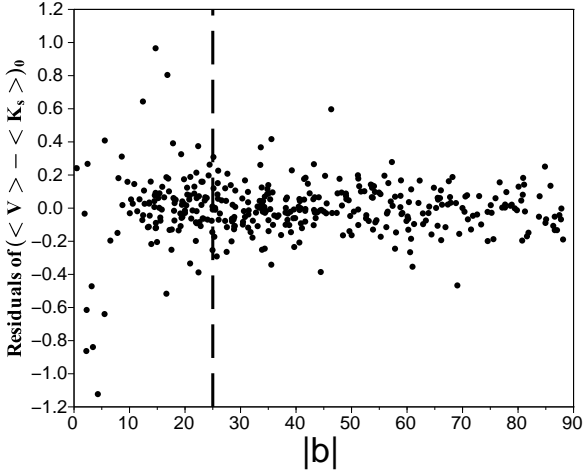}
  \caption{Plot of residuals of $(<V>-<K_s>)_0$ (equation~(\ref{abvkpar})) as a function of $|b|$. 
Note the sharp increase of residuals at low Galactic latitudes ($|b|~\leq~25^{\circ}$)}.
 \label{figresvk}
\end{figure}

Obviously, the 3D extinction model \citet{drimmel} should perform better for objects
at high Galactic latitudes and yield less accurate results in the vicinity of the
Galactic equator. This is immediately apparent from the plot of residuals of equation~(\ref{abvk})
as a function of $|b|$ (Fig.~(\ref{figresvk})). Residuals can be seen to increase sharply at $|b|~\leq~25^{\circ}$. We therefore
make a new least-squares fit leaving only stars with $|b|~>~25^{\circ}$:
\begin{eqnarray} \label{abvkpar1}
(<V>-<K_s>)_0  = (1.863 \pm 0.024)   \\  \nonumber
+ (0.144 \pm 0.016) \cdot [Fe/H] + 2.33 \cdot log P_F,  \\  \nonumber
\label{VKS}
\end{eqnarray}
with a scatter of $\sigma(<V>-<K_s>)_0$ = 0.11.
This result implies, given the adopted metallicity-$<M_V>$ relation (\ref{MVF}):
\begin{equation}
<M_{K_s}> = a_v - 1.863 + 0.088 (\pm0.026) \cdot [Fe/H] - 2.33 \cdot log P_F
\label{MKF}
\end{equation}
To make this relation as consistent as possible with the period-$<M_{K_s}>$ relation~(\ref{MKs}),
we set $a_v - 1.863 + 0.088 \cdot (<[Fe/H]>) =  -0.90$ (here $<[Fe/H]>$ = -1.419 is the average
metallicity of sample stars), implying $a_v$ = 1.088 and hence:
\begin{equation}
<M_V> = 1.088 + 0.232 \cdot [Fe/H]
\label{MVFF}
\end{equation}
and
\begin{equation}
<M_{K_s}> = -0.775 + 0.088(\pm0.026) \cdot [Fe/H] - 2.33 \cdot log P_F
\label{MKFF}
\end{equation}

We now assume that the WISE W1-band absolute magnitudes of RR Lyrae type variables depend both on metallicity
and fundamentalised period as:
\begin{equation}
<M_{W1}> = a_w + b_w \cdot [Fe/H] + c_w \cdot log P_F,
\label{MW}
\end{equation}
We subtract equation~(\ref{MW}) from equation~(\ref{MK}) to obtain (given that $(<K_s>-<W1>)_0 = <M_{K_s}> - <M_{W1}>$):

\begin{equation}
(<K_s>-<W1>)_0 = a_{kw} + b_{kw} \cdot [Fe/H] + c_{kw} \cdot log P_F,
\label{abkw}
\end{equation}
where $a_{kw} = a_k - a_w$, $b_{kw} = b_k - b_w$, and $c_{kw} = c_k - c_w$.
We then use the least squares method to solve equation set~(\ref{abkw}) for stars with $|b|~>~25^{\circ}$:
\begin{eqnarray} \label{abkwpar1}
(<K_s>-<W1>)_0  = (0.021 \pm 0.032)   \\  \nonumber
- (0.017 \pm 0.012) \cdot [Fe/H] - (0.039 \pm 0.071) \cdot log P_F,  \\  \nonumber
\end{eqnarray}
with a scatter of $\sigma(<K_s>-<W1>)_0$ = 0.07. Both the [Fe/H] and $log P_F$ terms are
statistically insignificant and we therefore seek the most simple solution setting
$b_{kw} = 0$ and $c_{kw}=0$ and leaving only the constant term:
\begin{equation}
(<K_s>-<W1>)_0  = (0.056 \pm 0.004)
\label{kwfin}
\end{equation}
with a scatter of $\sigma(<K_s>-<W1>)_0$ = 0.07
We thus derive, in view of  relation (\ref{MKFF}), the following
period-metallicity-$<M_{W1}>$ relation :
\begin{equation}
<M_{W1}> = -0.831 + 0.088(\pm0.026) \cdot [Fe/H] - 2.33 \cdot log P_F,
\label{MWF}
\end{equation} 
and the following $(<V>-<W1>)_0$ intrinsic colour calibration:
$$
(<V>-<W1>)_0 = 1.919(\pm0.024)
$$
\begin{equation}
 + 0.144(\pm0.016) \cdot [Fe/H] + 2.33 \cdot log P_F,
\label{VW0}
\end{equation} 
for RR Lyrae type stars. 
We now refine this calibration by deriving it directly from the observed $(<V>-<W1>)$ colours assuming that 
the period-metallicity-$<M_V>$ relation has zero log~$P_F$ slope and the metallicity slope equal to $b_v$=0.232$\pm$0.020,
and the  period-metallicity-$<M_{W1}>$ relation has the same log~$P_F$ slope as the   period-metallicity-$<M_{K_s}>$ relation 
(i.e., $c_w$=-2.33) and unknown metallicity slope:
$$
(<V>-<W1>)_0 = 1.884(\pm0.019)
$$
\begin{equation}
 + 0.120(\pm0.012) \cdot [Fe/H] + 2.33 \cdot log P_F
\label{VW1}
\end{equation} 
with a scatter of 0.086.
We now use this calibration to determine the colour excesses $E(<V>-<W1>) = (<V>-<W1>) - (<V>-<W1>)_0$
for most of the stars at low Galactic latitudes ($|b|~\leq~25$) and convert these colour excesses into $E(B-V)$ values
by the formula
$$
E(B-V) = E(<V>-<W1>) /(R_V-R_{W1}),
$$
where we adopt $R_V$ = 3.1 and $R_{W1}$ = 0.18 in accordance with
\citet{yuan}. No intensity-mean $<W1>$ magnitudes are available for four stars with $|b|~\leq~25$.
For these stars we use calibration~(\ref{VKS}) to compute the colour excesses 
$E(<V>-<K_s>) = (<V>-<K_s>) - (<V>-<K_s>)_0$, which we convert into $E(B-V)$ values
by the formula
$$
E(B-V) = E(<V>-<K_s>) / (R_V-R_{K_s}),
$$
where $R_{K_s}$ = 0.306 \citep{yuan}. 
For the remaining stars (i.e., those at $|b|~>~25$) we adopt the $E(B-V)$ values
$E(B-V) = A_V/R_V$ computed in terms of the 3D interstellar extinction  model
by \citet{drimmel} via the iterative procedure described above.

The  data used in this work (including the accurate coordinates, proper motions (with source and
standard-error estimate), radial velocities (with source and standard errors), intensity-mean $V$-,
$K_s$-, and $W1$-band magnitudes (with source and standard-error estimate), [Fe/H] (with source), 
period, and pulsation mode are listed in Table~\ref{mastertable} (the full  version  will be
available from the CDS). The columns of this table contain: (1) the GCVS name of the star; (2) and (3) - its J2000.0
equatorial coordinates $\alpha$ and $\delta$, respectively in decimal degrees; 
(4) the variability period in days; (5) the RR Lyrae type (AB or C);
(6) and (7) the intensity-mean $V$-band magnitude ($<V>$) and its standard error ($\sigma <V>$), respectively;
(8) the $V$-band interstellar extinction ($A_V$); (9) and (10) the metallicity [Fe/H] on the Zinn and West scale and
its reference code (Ref.), respectively; (11) - (13)  the intensity-mean $K_s$-band magnitude ($<K_s>$),
its standard error ($\sigma <K_s>$), and reference code (Ref.), respectively; 
(14)-(15)  the intensity-mean $W1$-band magnitude ($<W1>$) and its standard error ($\sigma <W1>$), respectively;
(16)-(18)   the average radial velocity ($V_R$), its standard error ($\sigma <V_R>$) (both in km/s), and reference code (Ref.), 
respectively; (19)-(20) the proper motion in right ascension ($\mu_{\alpha}$) and its standard error
($\sigma \mu_{\alpha}$) in mas/y, and (21)-22) the proper motion in right ascension ($\mu_{\delta}$) and its standard error
($\sigma \mu_{\delta}$) in mas/y.

\begin{table*}
   \tiny
  \centering
  \begin{minipage}{180mm}
  \caption{The first 10 lines of the catalogue of 
observational data for 403 Galactic field RR Lyrae variables. This is a sample of the full version, 
which is available in the online version of the article (see Supporting Information). }\label{mastertable}
\begin{tabular}{rccrrrrrrrrrr}
\hline
 (1) & (2)      &  (3)       &  (4)    & (5)  & (6)   &    (7)       &  (8)   &  (9)     & (10) & (11)    &  (12)          & (13) \\
Name & $\alpha$ &  $\delta$  &  Period & Type & $<V>$ & $\sigma <V>$ & $A_V$  & [Fe/H]   & Ref. & $<K_s>$ & $\sigma <K_s>$ & Ref. \\
     &  degrees &   degrees  &  (days) &      &       &              &        &          &      &         &                &      \\
\hline
SW    And & 005.929541 &  +29.401022  &  0.4423 & AB  &   9.712  &  0.009  &  0.113  &  -0.38   & 1  &  8.509  & 0.035 &  11  \\
XX    And & 019.364182 &  +38.950554  &  0.7229 & AB  &  10.687  &  0.009  &  0.004  &  -2.01   & 1  &  9.411  & 0.035 &  11  \\
XY    And & 021.676799 &  +34.068581  &  0.3988 & AB  &  13.680  &  0.019  &  0.137  &  -0.92   & 1  & 12.892  & 0.040 &  11  \\
ZZ    And & 012.395191 &  +27.022129  &  0.5546 & AB  &  13.082  &  0.018  &  0.125  &  -1.58   & 1  & 11.834  & 0.035 &  11  \\
AT    And & 355.628474 &  +43.014362  &  0.6170 & AB  &  10.694  &  0.018  &  0.414  &  -0.97   & 2  &  9.090  & 0.036 &  11  \\
BK    And & 353.775162 &  +41.102886  &  0.4216 & AB  &  12.970  &  0.037  &  0.344  &  -0.08   & 1  & 11.663  & 0.040 &  11  \\
CI    And & 028.784546 &  +43.765705  &  0.4848 & AB  &  12.244  &  0.018  &  0.211  &  -0.83   & 1  & 10.929  & 0.035 &  11  \\
DM    And & 353.003024 &  +35.196911  &  0.5916 & AB  &  11.943  &  0.037  &  0.439  &  -2.32   & 1  & 10.575  & 0.038 &  11  \\
DR    And & 016.294632 &  +34.218426  &  0.5328 & AB  &  12.431  &  0.037  &  0.122  &  -1.48   & 1  & 11.332  & 0.035 &  11  \\
NQ    And & 002.886174 &  +30.861319  &  0.6194 & AB  &  99.000  & 99.000  &  0.165  &  -1.43   & 8  & 14.668  & 0.091 &  12  \\
\hline
 &   & &  &   &  &   &   &   &  & & & \\
(Continued) &   & &  &   &  &   &   &   &  & & & \\
  & (14)  & (15) & (16) & (17)  & (18) & (19)  & (20)  & (21)  & (22)  & & & \\
Name &  $<W1>$  & $\sigma <W1>$ & $V_R$ & $\sigma V_R$  & Ref. & $\mu_{\alpha}$  & $\sigma \mu_{\alpha}$  & $\mu_{\delta}$  & $\sigma \mu_{\alpha}$ & & & \\
     &          &               &  (km/s)     & (km/s)              &      & (mas/yr)        & (mas/yr)               & (mas/yr)        & (mas/yr)              & & & \\
\hline
SW    And  &  8.464  &  0.009  &  -21.0 &   1.0  & 18 &    -7.6  &  1.1  &   -20.1  &  0.9 & &\\
XX    And  &  9.378  &  0.006  &    0.0 &   1.0  & 18 &    56.8  &  0.8  &   -35.0  &  0.7 & &\\
XY    And  & 12.617  &  0.005  &  -64.0 &  53.0  & 18 &    11.8  &  1.7  &    -8.3  &  2.4 & &\\
ZZ    And  & 11.795  &  0.010  &  -13.0 &  53.0  & 18 &    31.8  &  1.4  &   -15.3  &  1.6 & &\\
AT    And  &  9.012  &  0.004  & -228.0 &   1.0  & 18 &    -9.5  &  0.5  &   -50.5  &  0.7 & &\\
BK    And  & 11.614  &  0.008  &  -17.0 &   7.0  & 19 &     4.9  &  1.2  &    -1.2  &  1.2 & &\\
CI    And  & 10.978  &  0.006  &   24.0 &   5.0  & 18 &     0.8  &  0.9  &    -1.9  &  0.9 & &\\
DM    And  & 10.471  &  0.012  & -265.0 &  30.0  & 18 &     9.0  &  1.0  &     9.9  &  0.9 & &\\
DR    And  & 11.257  &  0.011  &  -81.0 &  30.0  & 18 &    28.2  &  1.1  &    -6.4  &  0.6 & &\\
NQ    And  & 14.850  &  0.044  & -242.0 &  15.0  & 20 &     5.4  &  4.7  &    -4.2  &  4.4 & &\\
\hline
\end{tabular}
\end{minipage}
\end{table*}

\normalsize

\section{Initial distances}

We now again deredden the observed
intensity-mean $<V>$, $<K_s>$, and $<W1>$ magnitudes using formulae~(\ref{DRV}--\ref{DRW})
and the finally adopted $E(B-V)$ values. We then determine the initial (assumed) distances to most of the RR Lyrae
stars of our sample based on the period-metallicity-$<M_{W1}>$ relation~(\ref{MWF}). We use the
the period-metallicity-$<M_{K_s}>$ relation~(\ref{MKFF}) to compute the 
distances to five stars with no intensity-mean $<W1>$ magnitudes.

Our working sample consists of all stars with simultaneously available $W1$-band intensity-mean magnitudes,
metallicities, radial velocities, and UCAC4 proper motions (a total of 387 stars).
Below we apply the method of statistical parallax to this sample to determine a
correction $\Delta <M_{W1}>$ to the zero point of relation~(\ref{MWF}) so that
\begin{equation}
<M_{W1(true)}> = <M_{W1}(adopted)> + \Delta <M_{W1}>.
\end{equation}

\section[]{THE METHOD}
Throughout this paper, we use the bimodal version of the classical maximum-likelihood 
statistical-parallax method \citep{m1} as described in Section~2.4 of \citet{dambis2}.
In this case the likelihood function of obtaining all observations (apparent magnitudes, fundamental periods,
metallicities, radial velocities, and proper motions of all stars of
the sample studied) depends on fourteen parameters: (1) three
components of the relative bulk motion of the objects of the
first ($U_{0(1)}$, $V_{0(1)}$, and $W_{0(1)}$) and second
($U_{0(2)}$, $V_{0(2)}$, and $W_{0(2)}$) subsamples with respect
to the Sun; (2) three diagonal components of velocity ellipsoid of
the first ($\sigma V_{rg(1)}^2$, $\sigma V_{\varphi(1)}^2$, and
$\sigma V_{\theta(1)}^2$) and second ($\sigma V_{rg(2)}^2$,
$\sigma V_{\varphi(2)}^2$, and $\sigma V_{\theta(2)}^2$)
subsamples, (3) the inverse distance-scale correction factor, $f$
(the same for both subsamples), and (4) the fraction $\alpha$ of
stars that belong to the first subsample (in the bimodal case
the fraction of stars that belong to the second subsample is evidently equal
to 1-$\alpha$).

\section[]{Results}

\citet{l3} and \citet{l1} showed that the
kinematic population of RR Lyraes in our Galaxy breaks
conspicuously into two subclasses: halo and thick-disc stars. 
As we pointed out in the previous section, we use
the model incorporating explicitly the bimodal velocity distribution 
and compute 14 unknown parameters
simultaneously. Table~\ref{res_all} lists the results based on the entire sample.

\begin{table*}
  \centering
  \begin{minipage}{140mm}
  \caption{Kinematical parameters and $W1$-band absolute-magnitude correction of
Galactic field RR Lyrae variables inferred from the entire sample (387 stars).  }\label{res_all}
\begin{tabular}{l  l  r  r r r r r r }
\hline
  Population & & Fraction of the sample  &  $U_0$ & $V_0$ & $W_0$ & $\sigma V_R$ & $\sigma V_{\varphi}$ & $\sigma V_{\theta}$ \\
 &   &  ($\alpha$ and 1-$\alpha$)   &   &   &   & km/s  &   &     \\
\hline
Halo & & 0.796 &  -6.8& -210.3 & -8.9 & 146.7 & 98.8 & 96.5  \\
        & &  $\pm$ 0.028   & $\pm$ 8.9 & $\pm$ 9.0 & $\pm$ 5.8 & $\pm$ 7.5 & $\pm$ 5.3 & $\pm$ 5.1  \\
     &   & &   &   &   &   &   &      \\
\hline
Disc &  & 0.204 & -11.3 & -38.9 & -15.6 & 45.2 & 37.0 & 26.6   \\
       & & $\pm$ 0.028   & $\pm$ 6.7 & $\pm$ 6.5 & $\pm$ 4.2 & $\pm$ 6.8 & $\pm$ 5.2 & $\pm$ 4.5     \\
   &        &      &      &      &    &     &    &      \\
\hline
   &   &   &   &   &   &   &   &          \\
 & &   &   &   & $\Delta <M_{W1}>$  & $=$  & +0.065  & $\pm$ 0.083     \\
\hline
\end{tabular}
\end{minipage}
\end{table*}

\begin{figure}
\includegraphics[width=0.5\textwidth]{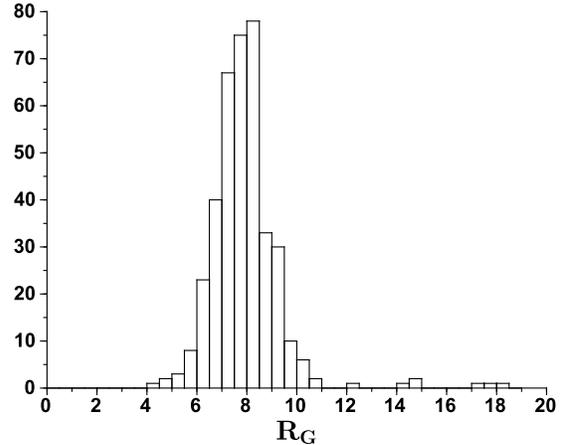}
  \caption{Histogram of Galactocentric distances of sample stars}.
 \label{rg_hist}
\end{figure}

Note, however, that despite its mathematical sophistication, our analysis has at least one weak point: it assumes that
all the quantities characterising the velocity distribution ( $U_0$, $V_0$, $W_0$, $\sigma V_R$, $\sigma V_{\theta}$, 
and $\sigma W$ for the halo and thick-disc subpopulations and their respective fractions) are location independent.
This is evidently not true, and the corresponding effect cannot be neglected given the large spread of Galactocentric distances 
of sample stars (Fig.~\ref{rg_hist}). To reduce the influence of this effect, we repeat our computations by applying
the method to a trimmed sample: we leave only stars with Galactocentric distances that differ no more than 1.6~kpc 
from the provisionally adopted Galactocentric distance of the Sun ($R_0$~=~8~kpc): $6.4~kpc~\leq~R_G~\leq~9.6~kpc$.
The corresponding results are presented in Table~\ref{res_trimmed}. Our result requires no 
absolute-magnitude zero point correction ($\Delta <M_{W1}>$  = +0.006   $\pm$ 0.088) and therefore we do not have
to  adjust the initial distances or intrinsic-colour calibrations~(\ref{abvkpar1}) and (\ref{kwfin}), which depend on the adopted 
extinction values, which, in turn, depend on the adopted distances via the 3D interstellar extinction model by \citet{drimmel}.

\begin{table*}
  \centering
  \begin{minipage}{140mm}
  \caption{Kinematical parameters and $W1$-band absolute-magnitude correction of
Galactic field RR Lyrae variables based on the bimodal solution
(336 stars with Galactocentric distances in the interval from 6.4 to 9.6~kpc).}\label{res_trimmed}
\begin{tabular}{l  l  r  r r r r r r }
\hline
  Population & & Fraction of the sample  &  $U_0$ & $V_0$ & $W_0$ & $\sigma V_R$ & $\sigma V_{\varphi}$ & $\sigma V_{\theta}$ \\
 &   &  ($\alpha$ and 1-$\alpha$)    &   &   &   & km/s  &   &     \\
\hline
Halo & & 0.777 &  -8.5& -214.0 & -9.3 & 153.3 & 100.4 & 96.8  \\
        & &  $\pm$ 0.029   & $\pm$ 9.4 & $\pm$ 9.6 & $\pm$ 6.2 & $\pm$ 8.5 & $\pm$ 5.8 & $\pm$ 5.4  \\
     &   & &   &   &   &   &   &      \\
\hline
Disc &  & 0.223 & -12.0 & -37.2 & -17.0 & 47.0 & 36.7 & 27.2   \\
       & & $\pm$ 0.029   & $\pm$ 7.0 & $\pm$ 6.1 & $\pm$ 4.3 & $\pm$ 6.9 & $\pm$ 5.1 & $\pm$ 4.4     \\
   &        &      &      &      &    &     &    &      \\
\hline
   &   &   &   &   &   &   &   &          \\
 & &   &   &   & $\Delta <M_{W1}>$  & $=$  & +0.006  & $\pm$ 0.088     \\
\hline
\end{tabular}
\end{minipage}
\end{table*}

To check our results for the eventual biases, we use the same procedure as
that employed by \citet{dambis2}. 
We generate for every our solution a set of 400 simulated samples 
consisting of objects with exactly the same sky locations  as those of stars in
the real sample. We simulate the "observed" parameters of each object (i.e., the 
distance, radial velocity, and proper-motion components) based on the 
distance of the corresponding star (which we divide by the inferred distance-scale correction
factor $f$ plus the cosmic scatter), the mean velocity components (($U_0$, $V_0$, $W_0$) of one of the two subpopulations
(chosen randomly in accordance with the inferred fractions $\alpha$ and 1-$\alpha$) with
the corresponding velocity ellipsoid (defined by $\sigma V_{rg}$, $\sigma V_{\varphi}$, and $\sigma V_{\theta}$) 
superimposed. We finally add to the radial velocities and proper-motion components so computed
the corresponding  simulated random errors with zero means and variances determined by the 
standard errors of the actually observed quantities. We  apply our bimodal method to each of the simulated samples
to estimate the mean offsets of all the inferred parameters. 

We find our method to yield unbiased estimates for all the parameters in the case considered, and therefore
apply no corrections to our results. We adopt the parameters given by the solution based on the trimmed sample
(Table\ref{res_trimmed})as our final result, and, in particular, the corresponding corrected zero point for the infrared
PL relation, which is is 0.026$^m$ fainter than that the initially adopted value (see formula~(\ref{MWF})), 
implying the following corrected W1-band period-metallicity-luminosity relation:
$$
<M_{W1}> = -0.825(\pm0.088)
$$
\begin{equation}
 + 0.088(\pm0.026) \cdot [Fe/H] - 2.33 \cdot log P_F.
\label{MWC}
\end{equation} 
We also apply the same zero-point correction to the V-band metallicity-luminosity relation given by formula~(\ref{MVFF}) and
the $K_s$-band period-metallicity-luminosity relation given by formula~(\ref{MKFF}) to obtain the following corrected
relations:
\begin{equation}
<M_V> = 1.094(\pm0.091) + 0.232(\pm0.020) \cdot [Fe/H]
\label{MVFC}
\end{equation}
and
$$
<M_{K_s}> = -0.769(\pm0.088)
$$
\begin{equation}
 + 0.088(\pm0.026) \cdot [Fe/H] - 2.33 \cdot log P_F.
\label{MKFC}
\end{equation}
The implied period-metallicity-luminosity relation for the CIT K band then is:
$$
<M_{K_{CIT}}> = -0.750(\pm0.088) +
$$
\begin{equation}
 0.088(\pm0.026) \cdot [Fe/H] - 2.33 \cdot log P_F.
\label{MKFCIT}
\end{equation}

\subsection{Comparison to earlier work}

The only previous WISE W1-band luminosity calibration of RR Lyrae type variables is that of \citet{klein},
who found $<M_{W1}> = (– 0.421 \pm 0.014) –- (1.681 \pm 0.147) \cdot log (P/0.50118 day)$ with no evidence for
metallicity term by computing  posterior distances of 76 well observed RR Lyrae based on
the optically constructed prior distances. Our PL relation is $\sim$~0.16$^m$ fainter at $P = 0.50118$ and
[Fe/H]=-1.5, and this is due entirely to the prior distance estimates adopted by the above authors, which are 
based on the $<M_V>$ - [Fe/H] relation $<M_V>$=0.93 + 0.23$\cdot$[Fe/H] \citep{chaboyer}. The difference in the
slope appears to be due to fact that \citet{klein} do not fundamentalise the periods of RRc type stars. We
believe this to be a questionable approach given that (a) RRc type stars form a well-defined 
period-shifted (by $\delta$~log($P$)=-0.127) branch of the $K$-band PL relation (see, e.g., \citet{fs})
and (b) intrinsic $K_s$-W1 colours are practically constant for A-F type stars, implying that $<M_{W1}>$ may serve
as a proxy for $<M_{Ks}>$ and that the period-$<M_{W}>$ relation should have practically the same slope as
the period-$<M_{K_s}>$ relation (i.e., -2.33).

Our RR Lyrae $V$-band luminosity scale zero point ($<M_V>$=+0.72$\pm$0.09 at [Fe/H]=-1.6),
albeit consistent with most of the analyses based on the method of statistical parallax, 
differs markedly from the most recent
such study of c-type RR Lyrae variables by \citet{k12}: they found $<M_{V, RRc}>$~=~+0.52~$\pm$~0.11 at
a metallicity of [Fe/H]~=~-1.59, whereas formula~(\ref{MVFC}) implies $<M_{V}>$~=~+0.73$\pm$0.09 for the same metallicity.
This can be compared with the statistical-parallax based estimates $<M_{V}>$~=~+0.71$\pm$0.12  \citep{l1} and
$<M_{V}>$~=~+0.75$\pm$0.13 \citep{pg1}, both at   [Fe/H]=-1.61; $<M_{V}>$~=~+0.73$\pm$0.14  \citep{f1} at [Fe/H]=-1.53;
$<M_{V}>$~=~+0.69$\pm$0.10  \citep{ts} at   [Fe/H]=-1.58; $<M_{V}>$~=~+0.65$\pm$0.23  \citep{luri} at   [Fe/H]=-1.51, and
$<M_{V}>$~=~+0.76--0.78$\pm$0.10  \citep{dr1} at   [Fe/H]=-1.60.

Our zero point of the $K$-band luminosity calibration is also consistent with the previous statistical-parallax
determinations
$<M_{K_s}>$ = -0.82$\pm$0.12 -2.33$\cdot$log($P_F$) \citep{d1},
$<M_{K_s}>$ = -0.89$\pm$0.09 -2.33$\cdot$log($P_F$) \citep{dv}, 
$<M_{K_s}>$ = -0.94$\pm$0.06 -2.33$\cdot$log($P_F$) \citep{rdz}, 
and
$<M_{K_s}>$ = -0.818$\pm$0.08 -2.33$\cdot$log($P_F$) \citep{dambis2} compared to
$<M_{K_s}>$ = -0.901$\pm$0.09 -2.33$\cdot$log($P_F$) (formula~(\ref{MKFC}) with [Fe/H] fixed at -1.5).

Both our $V$- and $K_s$-band absolute-magnitude calibrations agree well with the results obtained
using the Baade-Wesselink method:  $<M_{K}>$ = -0.88 -2.33$\cdot$log($P_F$) and 
$<M_V>$ = 1.02+0.16$\cdot$[Fe/H] (the latter implies $<M_{V}>$~=~+0.76$\pm$0.03 at [Fe/H] = -1.6) \citep{j1} and
$<M_V>$ = 0.98+0.20$\cdot$[Fe/H] (implying $<M_{V}>$~=~+0.66$\pm$0.05 at [Fe/H] = -1.6) \citep{f2}).

At the same time, our RR Lyrae luminosity scale is inconsistent with HST trigonometric parallaxes 
of five RR Lyrae type variables: our $<M_V>$, $<M_{Ks}>$, and $<M_{W1}>$ zero points are
0.34$\pm$0.05, 0.32$\pm$0.05, and 0.34$\pm$0.05 fainter than those inferred by applying 
the reduced parallax method  to the HST parallax data \cite{b3}. The reason for such a 
discrepancy remains unclear. 

\begin{figure}
\includegraphics[width=0.5\textwidth]{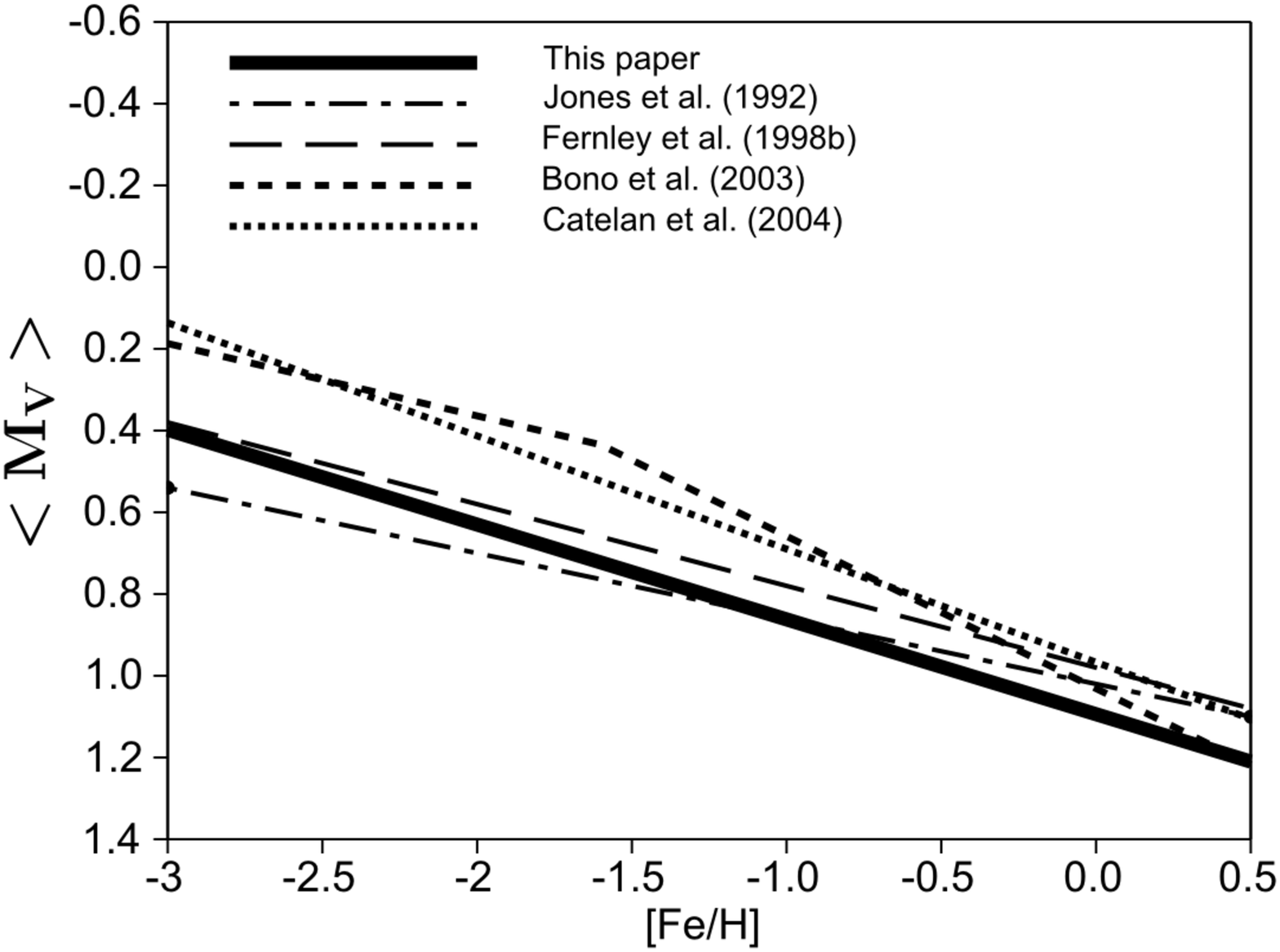}
  \caption{Comparison of our metallicity-$<M_V>$ relation (thick solid line) with published theoretical relations
\citep{bono, catelan} and relations based on
the Baade--Wesselink method \citep{j1,f2}}.
 \label{comp_mv}
\end{figure}

\begin{figure}
\includegraphics[width=0.5\textwidth]{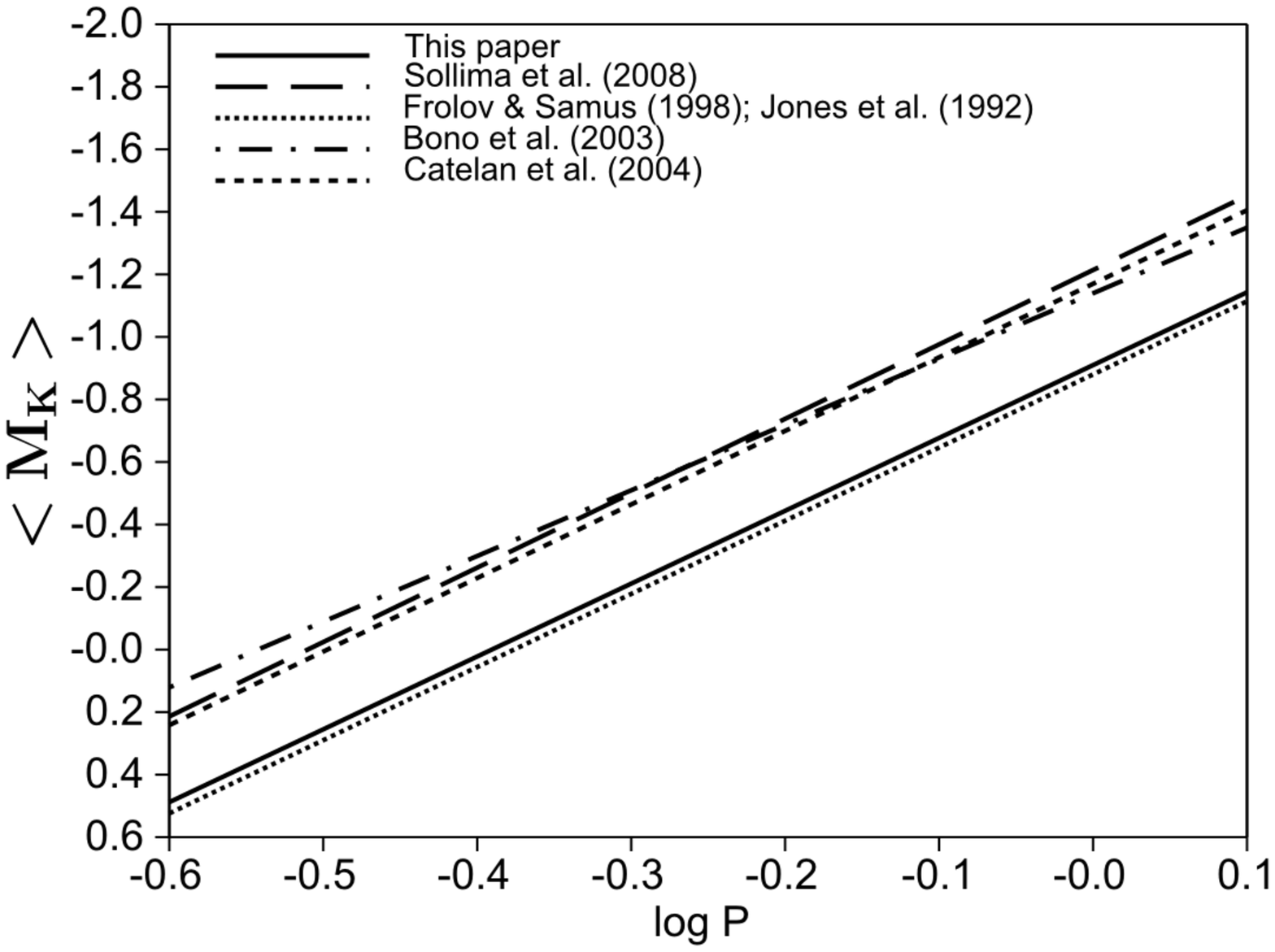}
  \caption{Comparison of our period-$<M_K>$ relation (solid line) with published theoretical relations
\citep{bono, catelan}, a relation based on the Baade--Wesselink method \citep{j1}, and two relations based
on analyses of cluster RR Lyrae type stars \citep{fs, sollima}}.
 \label{comp_mk}
\end{figure}

Our $V$- and $K$- band luminosity scales are also appreciably fainter than the estimates based on analyses  based on 
theoretical models \citep{bono, catelan} and that defined by the period-metallicity-$<M_K>$ relation
of \citet{sollima}: $<M_K>$=-1.07 - 2.38$\cdot$log $P_F$ +0.09$\cdot [Fe/H]_{ZW}$.  The latter is tied to the 
HST parallax of RR Lyrae itself  and is therefore naturally $\sim$0.3$^m$ brighter than our period-metallicity-$<M_K>$
relation (see above). Note, however, that our metallicity slope, which is equal to +0.088$\pm$0.026, practically coincides with that 
of the relation derived by \citet{sollima}, but is appreciably shallower than the values of
0.231$\pm$0.012 and 0.277 estimates derived theoretically by \citet{bono} and \citet{catelan}, respectively. We compare our
metallicity-$<M_V>$ relation with the published ones in Fig~\ref{comp_mv}. A similar comparison of period-$<M_K>$ relations 
(at [Fe/H]=-1.6) is made in Fig.~\ref{comp_mk}.

We thus see that, as always before, the method of statistical parallax, like the Baade--Wesselink method, 
continues to yield RR Lyrae luminosities and distances that are at the lower end of the estimates obtained 
using various techniques. The cause of the discrepancies with other methods remains unknown and can hardly be attributed
to photometric or astrometric errors given the dramatic improvement of the quality of the corresponding data
both in terms of random errors and homogeneity. Neither can it be due to eventual systematic errors in the
estimated amount of interstellar extinction, which is rather small for most of the RR Lyrae variables considered:
its average value in the $V$ band  is less than 0.2$^m$, and becomes negligible in the $K_s$ and $W1$ bands (0.019$^m$ and 
0.013$^m$, respectively). Furthermore, the $<M_{K_s}>$- and $<M_{W1}>$-band absolute magnitudes depend only 
slightly on metallicity and therefore metallicity errors are very unlikely to introduce any significant biases into
these quantities. As shown by \citet{pg1, pg2}, the same is true for eventual radial-velocity errors and errors
in the adopted kinematical model. We hope that the upcoming GAIA mission will finally resolve this controversy.

The kinematical results obtained appear to be consistent with earlier statistical-parallax based results 
for RR Lyrae type stars (\citep{l1, f1, dr1, dambis2, k12}) and with those of a recent analysis of  
the kinematics of BHB stars by halo stars by \citet{kafle} (see Fig.~3 in their paper) and 
the fundamental kinematical study  of halo stars by \citet{carollo} (Supplemental Table~1),
\citet{carollo10}. 
As in our previous paper \citep{dambis2}, we find no signs of the outer halo 
population with retrograde rotation and
large $\sigma V_R$ and $\sigma V_{\theta}$ velocity-dispersion components, which was to be expected
given that our sample contains very few distant enough stars. 

\section{Distance-scale implications}

\citet{c2} estimated the distance to the Galactic centre to be $R_0$~=~7.80~$\pm$~0.40~kpc  based on the 
$K_{CIT}$-band photometry of Galactic bulge RR Lyraes and the assumed PL relation by \citet{j1} (formula~(\ref{MKJ})).
\citet{pietrukowicz} find the average metallicity of Galactic bulge RR Lyrae type variables to be
$<[Fe/H]>$~=~-1.02 on the scale of \citet{jurcsik}. This corresponds to $<[Fe/H]>_{ZW}$ = 1.028~$\cdot$~(-1.02) - 0.242 =
-1.29 \citep{papadakis} on the \citet{zw} metallicity scale. The resulting zero point of the (CIT) K-band PL relation 
for bulge RR Lyraes is then equal to -0.750 + 0.088$\cdot$(-1.29)=-0.863 (formula~(\ref{MKFCIT})),
implying a solar Galactocentric distance of  
$R_0$~=~(7.80~$\pm$~0.40)~$\cdot$~10$^{-0.2\cdot(-0.863 - (-0.88))}$~=~7.74~$\pm$~0.40~kpc. \citet{groenewegen}
performed $K_s$-band observations of 39 type-II Cepheids and 37 RR Lyrae type stars in the Galactic bulge and estimated the 
RR Lyrae based distance to the Galactic centre to be $R_0$~=~7.87~$\pm$~0.64~kpc~$\pm$~0.26~kpc using the
period-metallicity-$<M_K>$ relation by \citet{sollima1}. When recalibrated to our period-metallicity-$<M_K>$ relation, their
RR Lyrae type star photometry yields a solar Galactocentric distance of $R_0$~=~7.19~$\pm$~0.40~kpc~$\pm$~0.24~kpc.
Based on OGLE II observations of stars in the Galactic bulge,
\citet{csf} find the average extinction-corrected intensity-mean $V$-band magnitude of RR Lyrae type stars
in the Baade window to be $<V_0>$=15.44$\pm$0.05. Given the average metallicity of [Fe/H]=-1.0 of bulge stars,
our [Fe/H]-$<M_V>$ relation~(\ref{MVFC}) yields an absolute magnitude of $<M_V>$=+0.862, implying a GC distance
modulus of $DM_{GC}$ = 14.58$\pm$0.10, or $R_0$=8.24$\pm$0.39~kpc. The average of the three estimates,
$R_0$=7.73$\pm$0.36~kpc, is consistent both with the result 
based on the PL relation of type II Cepheids ($R_0$ = 7.64 $\pm$ 0.21~kpc
\citep{flkvw} and $R_0$ = 7.99 $\pm$ 0.09~kpc \citep{groenewegen}) and with the direct estimate based on the  orbital solution for the 
star S0-2 orbiting the supermassive black hole at the Galactic centre ($R_0$ = 8.28 $\pm$ 0.15 $\pm$ 0.29~kpc
\citep{gillessen}).

If applied to the $K_s$-band photometry for 30 RR Lyrae type
variables in Reticulum \cite{dallora} ([Fe/H]=-1.71),  PML relation
(\ref{MKFC}) yields a distance
modulus of 18.22 $\pm$ 0.01 $\pm$ 0.09 ($D$ = 44.1 $\pm$ 0.2 $\pm$ 1.7 kpc) for this cluster, which is
purportedly associated with the LMC.  The $D_{Reticulum}/D_{LMC centre}$ ratio can be estimated using a geometric model 
of the LMC, which is parametrised by the inclination $i$ and position angle of the line of nodes, $PA$ by the
following formula:
$$
D_{Reticulum}/D_{LMC centre}=
$$
\begin{equation}
cos~i/[cos~i~cos~\rho - sin~i~sin~\rho~sin(\Phi-PA)]
\end{equation}
\citep{vandermarel}, where $\rho$=11.4$^{\circ}$ and $\Phi$=329$^{\circ}$ are the angular distance and position
angle of Reticulum with respect to the LMC, respectively \citep{schommer}. Recent estimates of the LMC orientation 
parameters $PA$ and $i$ \citep{vandermarel, olsen, nikolaev, persson, koerwer, subramanian, subramanian2} yield
$D_{Reticulum}/D_{LMC centre}$ ratios ranging from 0.959 to 1.041 corresponding to distance-scale corrections  ranging from
-0.09 to +0.09 with a mean of -0.01 and a standard deviation of 0.06. Hence the Reticulum distance derived yields 
an LMC distance modulus of 18.21 $\pm$ 0.01 $\pm$ 0.09 $\pm$ 0.06 ($D$ = 43.9 $\pm$ 0.2 $\pm$ 1.7 kpc $\pm$ 1.2 kpc).
The $K_s$-band photometry
of 10 RR Lyrae type stars in the inner regions of the LMC with 
spectroscopically measured [Fe/H] \citep{borissova}
yields a distance modulus of 18.38 $\pm$ 0.04 $\pm$
0.09 ($D$ = 47.4 $\pm$ 0.9 $\pm$ 2.0 kpc), and the similar data (65 stars) obtained by \citet{sz} imply an
LMC distance modulus of 18.30 $\pm$ 0.02 $\pm$ 0.09 ($D$ = 45.7 $\pm$ 0.7 $\pm$ 1.9 kpc). 
The $V$-band intensity-mean magnitudes and metallicities of 89 RR Lyrae type stars determined by 
\citet{gratton} analyzed in terms of our metallicity-$V$-band luminosity relation~(\ref{MVFC}) yield
an LMC distance modulus of 18.34 $\pm$ 0.02 $\pm$ 0.09 ($D$ = 46.6 $\pm$ 0.7 $\pm$ 2.1 kpc).
We thus derive an average 
LMC distance modulus estimate of  DM$_{LMC}$ = 18.32 $\pm$ 0.09 ($D$ = 46.1 $\pm$ 2.0 kpc).
This result is consistent with the estimate based on the PL relation 
of classical Cepheids and their HST trigonometric parallaxes: \citet{flkvw} report $DM_{LMC} = 18.39 \pm 0.05$ as
inferred from the estimate of \citet{b2} with the metallicity correction of
\citet{macri}. Our LMC distance 
modulus also agrees with the recent LMC distance estimate based on type II Cepheids ($DM_{LMC} = 18.37 \pm 0.09$ 
\citep{flkvw}). However, it is $\sim$1.3-2.2$\sigma$ smaller than the 
DM$_{LMC}$ = 18.46 $\pm$ 0.06 -- 18.61 $\pm$ 0.05 ($D$ = 49.2 $\pm$ 1.4 kpc -- 52.7 $\pm$ 1.2 kpc) estimates
implied by $V$- and $K_s$-band photometry of RR Lyrae type variables in the LMC and HST triginometric parallaxes 
of five Galactic RR Lyraes \citep{b3}, 
and $\sim$1.5$\sigma$ smaller than the DM$_{LMC}$ = 18.49 $\pm$ 0.05 ($D$ = 49.97 $\pm$ 1.11 kpc)
estimate inferred from observations of eclipsing binaries \citep{pietrzynski} and the 
DM$_{LMC}$ = 18.48 $\pm$ 0.04 ($D$ = 49.66 $\pm$ 0.92 kpc) estimate inferred from 
3.6- and 4.5-$\mu$m observations of classical Cepheids made by Spitzer Space Telescope and calibrated
by HST guide-star parallaxes of 10 Galactic Cepheids \citep{monson}.

\citet{clementini} analysed archival HST photometry of stars in the metal-poor globular cluster B514 in the M31 galaxy
and identified 89 RR Lyrae type variables. Our $V$-band luminosity calibration (formula~(\ref{MVFC})) implies
a distance modulus of $DM_{B514}$ = 24.32 $\pm$ 0.12 ($D$ = 731 $\pm$ 41 kpc)

\citet{federici} collected and analysed archival HST $BVI$ photometry for 48 globular clusters in the M31 galaxy.
They determined the apparent horizontal-branch $V$-band magnitudes, $M_V(HB)$, and 
estimated the reddenings and metallicities of these clusters by comparing their observed 
colour-magnitude diagrams to the colour-magnitude diagram ridge lines of a set of Galactic reference globular clusters.
Given  that in a globular cluster $M_V(HB)$ = $<M_V(RR Lyr)>$, we apply formula~(\ref{MVFC}) to
the data of \citet{federici} to obtain an M31 distance modulus estimate of 
$DM_{M31}$ = 24.201 $\pm$ 0.014 $\pm$ 0.090 ($D$ = 692 $\pm$ 5 $\pm$ 28 kpc).

When  applied to the results of an an extensive survey of RR Lyrae type variables in three fields along the
major axis of the M33 galaxy by \citet{yang}, our $V$-band metallicity-luminosity relation~(\ref{MVFC}) yields
a distance modulus of  $DM_{M33}$ = 24.36 $\pm$  0.09 ($D$ = 745 $\pm$ 31 kpc).

Table~\ref{LG} summarises the RR Lyr (or horizontal-branch) based distance 
estimates mentioned above,  and other such distance estimates for a number of  Local-Group galaxies. Our distance
moduli are, on the average, 0.13 smaller than most recent published estimates. This difference is due 
mostly to the difference in the zero points of the underlying calibrations, which are generally based on some reference
LMC distance modulus (cf. our  $DM_{LMC}$ = 18.34$\pm$0.09 with the estimates $DM_{LMC}(Freedman)$ = 18.477$\pm$0.033
and $DM_{LMC}(Riess)$ = 18.486$\pm$0.065 adopted by \citet{freedman} and \citet{riess}, respectively.

\begin{table*}
  \centering
  \begin{minipage}{150mm}
  \caption{Distance to Local Group galaxies estimated from the data for RR Lyr type variables (RR) 
or horizontal-branch stars (HB) using relations~(\ref{MVFC}) (for $V$-band data) or (\ref{MKFC}) (for $K$-band data).
Published distance moduli are representative only to give an idea of (generally) most recent previous estimates.}\label{LG}
\begin{tabular}{l  l  r  l r r l r r r }
\hline
  Galaxy  & Objects & Filter  & Instrument & Distance        & Distance,                 & Data  & Rem. & Published & Ref. \\ 
          &         &         &            & modulus         & kpc                       & ref.  &      & DM        &      \\
\hline
Milky Way & RR      &   $K$   & 4-m CTIO   & 14.44$\pm$0.11  & 7.74$\pm$0.40             & 1     &      &           &      \\
Milky Way & RR      &   $K_s$ & ESO NTT    & 14.29$\pm$0.11  & 7.19$\pm$0.37             & 1a    &      &           &      \\
(Galactic & RR      &   $V$   & OGLE-II    & 14.58$\pm$0.10  & 8.24$\pm$0.39             & 3     &      &           &      \\
centre)   &         &          & {\bf Mean:}&{\bf 14.44$\pm$0.10}  & {\bf 7.73$\pm$0.36} &       &      & {\bf 14.59$\pm$0.09}     & 31     \\
\hline
  LMC     & RR      &   $K_s$ & ESO VLT,   & 18.38$\pm$0.10  & 47.4$\pm$2.2              & 4     &      &           &      \\
          &         &         & ESO NTT    &                 &                           &       &      &           &      \\
  LMC     & RR      &   $K_s$ & ESO NTT    & 18.30$\pm$0.09  & 45.7$\pm$1.9              & 5     &      &           &      \\
  LMC     & RR      &   $K_s$ & ESO NTT    & 18.21$\pm$0.11  & 43.9$\pm$2.1              & 6     & a    &           &      \\
  LMC     & RR      &   $V$   & ESO VLT    & 18.34$\pm$0.09  & 46.6$\pm$2.1              & 7     &      &           &      \\
          &         &         & {\bf Mean:}&{\bf 18.32$\pm$0.09}  & {\bf 46.1$\pm$2.0}   &       &      & {\bf 18.477$\pm$0.033}   & 26     \\
\hline
  SMC     & RR      &   $V$   & OGLE-III   & 18.72$\pm$0.09  & 55.5$\pm$2.3              & 8     &      &           &      \\
  SMC     & RR      &   $K_s$ & ESO NTT    & 18.71$\pm$0.17  & 55.2$\pm$3.5              & 9     &      &           &      \\
          &         &         & {\bf Mean:}&{\bf 18.72$\pm$0.09}  & {\bf 55.5$\pm$2.3}   &       &      & {\bf 18.93$\pm$0.10}    &  27    \\
\hline
  M31     & RR      &   $V$   & HST        & 24.32$\pm$0.12  & 731$\pm$41                & 10     & b    &           &      \\
  M31     & HB      &   $V$   & HST        & 24.20$\pm$0.09  & 692$\pm$28                & 11    &      &           &      \\
          &         &         & {\bf Mean:}&{\bf 24.24$\pm$0.09}  & {\bf 705$\pm$30}     &       &      & {\bf 24.38$\pm$0.06}  &   23   \\
\\
          &         &         & {\bf M31}  & {\bf satellites}&                           &       &      &           &      \\
  M32     & RR      &   $V$   & HST        & 24.26$\pm$0.12  & 711$\pm$40                & 12    &      & {\bf 24.51$\pm$0.14}  &  28    \\
  NGC147  & RR      &   $V$   & HST        & 24.00$\pm$0.16  & 631$\pm$48                & 13    &      & {\bf 24.26$\pm$0.06}  &  29    \\
  And I   & RR      &   $V$   & HST        & 24.23$\pm$0.11  & 701$\pm$36                & 14    &      & {\bf 24.31$\pm$0.05}  &  29    \\
  And II  & RR      &   $V$   & HST        & 23.93$\pm$0.11  & 611$\pm$32                & 15    &      & {\bf 24.00$\pm$0.05}  &  29   \\
  And III & RR      &   $V$   & HST        & 24.18$\pm$0.11  & 685$\pm$35                & 14    &      & {\bf 24.30$\pm$0.06}  &  29   \\
  And V   & RR      &   $V$   & HST        & 24.41$\pm$0.09  & 762$\pm$32                & 16    &      & {\bf 24.35$\pm$0.06}  &  29   \\
  And VI  & RR      &   $V$   & HST        & 24.37$\pm$0.09  & 748$\pm$32                & 17    &      & {\bf 24.47$\pm$0.07}  &  30   \\
  And XI  & RR      &   $V$   & HST        & 24.17$\pm$0.09  & 682$\pm$29                & 18    &      & {\bf 24.33$\pm$0.05}  &  18   \\
  And XIII& RR      &   $V$   & HST        & 24.46$\pm$0.09  & 780$\pm$33                & 18    &      & {\bf 24.62$\pm$0.07}  &  18   \\
\hline
  M33     & RR      &   $V$   & HST        & 24.36$\pm$0.09  & 745$\pm$31                & 19    &      & {\bf 24.53$\pm$0.11}          & 24     \\
\hline
  IC1613  & RR      &   $V$   & HST        & 24.19$\pm$0.09  & 689$\pm$29                & 20    &      & {\bf 24.34$\pm$0.03}          & 25     \\
\hline
  Sex dSph& RR      &   $V$   & 3.6-m CFH  & 19.74$\pm$0.21  & 88.7$\pm$8.9              & 21    &      & {\bf 19.90$\pm$0.06}           & 21     \\
\hline
  Cetus   & RR      &   $V$   & HST        & 24.37$\pm$0.09  & 748$\pm$32                & 22    &      & {\bf 24.46$\pm$0.12}      & 22     \\
\hline
  Tucana  & RR      &   $V$   & HST        & 24.66$\pm$0.09  & 855$\pm$36                & 22    &      & {\bf 24.74$\pm$0.12}      & 22     \\
\hline
\end{tabular}
\begin{tabular}{l}
Data sources: 1. \citet{c2}; 2. \citet{groenewegen}; 3. \citet{csf}; 4. \citet{borissova}; \\
5. \citet{sz}; 6. \citet{dallora}; 7. \citet{gratton}; 8. \citet{kh}; \\
9. \citet{sz2}; 10. \citet{clementini}; 11. \citet{federici}; 12. \citet{sarajedini}; \\
13. \citet{ys2}; 14. \citet{pritzl}; 15. \citet{pritzl2}; 16. \citet{ms}; \\
17. \citet{pritzl3}; 18. \citet{ys}; 19. \citet{yang}; 20. \citet{bernard}; \\
21. \citet{lee}; 22. \citet{bernard2}; 23. \citet{rfv} (classical Cepheids);\\
24. \citet{sbmw} (classical Cepheids); 25. \citet{trs} (classical Cepheids); \\
26. \citet{freedman} (classical Cepheids); 27. \citet{inno}; 28. \citet{fiorentino}; \\
29. \citet{conn} (tip of the red-giant branch); 30. \citet{mcc} (tip of the red-giant branch);\\
31. \citet{gillessen}.\\
Remarks: {\bf a}. Reticulum globular cluster. It may be rather distant from the centre of the LMC\\
and we therefore use a geometric model to estimate the distance to the LMC centre (see text). {\bf b}. B514 globular cluster.  \\
\end{tabular}
\end{minipage}
\end{table*}

\section{Rotation of the RR Lyrae populations}
Our estimate for the distance to the Galactic center, $R_0$=7.73$\pm$0.36~kpc, combined with
the proper motion of Sgr A$^*$ along Galactic longitude, 
$\mu_{SgrA^{*}}$ = 6.379~$\pm$~0.026 mas/yr \citep{rb}, implies a solar velocity of $V_{\cdot}$=234$\pm$10~km/s.
It follows from this that the halo RR Lyrae population exhibits marginal prograde rotation with a velocity
of 20$\pm$15~km/s. The thick-disc RR Lyrae population rotates at a velocity of 193$\pm$13~km/s.

\section{Cosmology}
Our result, if taken at face value, would upscale  the Hubble 
constant as estimated by \citet{freedman} ($H_0$ = 74.3 $\pm$ 2.1 km/s/Mpc) to
$H_0$ = 80.0 $\pm$ 3.4 km/s/Mpc, and imply, if combined with $\Omega_\Lambda$= 0.74$\pm$0.02 \citep{mr}
based on an analysis of a sample of 580 type Ia supernovae in terms of a flat, LCDM cosmological model, an expansion age of 
12.5~Gyr, which is consistent with the globular-cluster ages of 12--13~Gyr  \citep{carretta}.

\section{Conclusions}
We collected homogenized data for 403 local Galactic field RR Lyrae variables: intensity-mean $V$-band, 2MASS $K_s$-band, and 
WISE $W1$-band magnitudes for 384, 403, and  398 stars, respectively; metallicity estimates on the \cite{zw} scale for 402 stars,
UCAC4 proper motions for 393 stars, as well as radial velocities and interstellar-extinction estimates for all 403 stars. We
used the photometric data combined the 3D interstellar extinction model by \citet{drimmel} to calibrate RR Lyrae star colours
in terms of fundamental period and metallicity and used these calibrations to determine the log $P_F$ and metallicity slopes
of the 2MASS $K_s$-band and WISE $W1$-band infrared period-metallicity-luminosity relations.
We performed a bimodal statistical-parallax solution on the subsample of stars with both WISE $W1$-band intensity-mean magnitudes
and UCAC4 proper motions available. We found the velocity components (relative to the Sun) of the halo and thick-disc populations along the
direction of Galactic rotation to be $V_0(Halo)$=-214$\pm$10 km/s and $V_0(Disc)$=-37$\pm$6 km/s, respectively. The inferred velocity-ellipsoid
components of are ($\sigma V_R(Halo), \sigma V_{\phi}(Halo), \sigma V_{\theta}(Halo)$) =
$(153 \pm 9, 101 \pm 6, 96 \pm 5)$ km s$^{-1}$ and  
($\sigma V_R(Disc), \sigma V_{\theta}(Disc), \sigma W(Disc)$) =
$(46 \pm 7, 37 \pm 5, 27 \pm 4)$ km s$^{-1}$ for the halo and thick-disc, respectively, and the fraction of
thick-disc stars, 0.22 $\pm$ 0.03. Our kinematical results agree well with those of other studies, and
our final  IR period-metallicity-luminosity relations and the $V$-band metallicity-luminosity relation are consistent
with most of the previous statistical-parallax based estimates (with the exception of the statistical-parallax analysis
of c-type RR Lyrae variables by \citet{k12}) and the results of the application of the Baade--Wesselink method, 
but generally fainter than implied by the results obtained using other methods. The distance moduli of Local Group galaxies estimated using
our derived IR and optical luminosity calibrations are, on the average, 0.13 smaller than recently published
determinations, implying a Hubble constant of $\sim$ 80 km/s and a rather small -- 12.5~Gyr--expansion age
of the Universe.

\section{Acknowledgements}
We thank the reviewer, Prof. G. Bono, for his valuable comments, which improved the
final version of the paper.  This publication makes use of data
products from the Two Micron All Sky Survey, which is a joint
project of the University of Massachusetts and the Infrared
Processing and Analysis Center/California Institute of
Technology, funded by the National Aeronautics and Space
Administration and the National Science Foundation,
and from the Wide-field Infrared Survey Explorer, which is a 
joint project of the University of California, Los Angeles, 
and the Jet Propulsion Laboratory/California Institute of 
Technology, funded by the National Aeronautics and Space Administration. 
This research has made use of NASA's Astrophysics Data System.
This work is supported by the Russian Foundation for Basic Research
(projects nos.~13-02-00203-a and 11-02-00608-a). AYK and RS acknowledge the support
from the National Research Foundation (NRF) of South Africa

\end{document}